\documentclass[
a4paper,
showpacs,
twocolumn,
superscriptaddress,reprint,amsmath,amssymb,aps,prb,longbibliography,showpacks]{revtex4-2}
\usepackage{amssymb}
\usepackage{amsmath}
\usepackage{amsfonts}
\usepackage{braket}
\usepackage{graphicx}
\usepackage{bm}
\usepackage{color}
\usepackage{multirow}
\usepackage{natbib}
\usepackage{hyperref}
\usepackage[normalem]{ulem}
\usepackage{verbatim}
\usepackage{dcolumn}
\usepackage{ulem}
\usepackage{float}

\usepackage[dvipsnames]{xcolor}

 \def \be {\begin{equation}}
\def \ee {\end{equation}}
\def \bea {\begin{align}}
\def \eea {\end{align}}
\def \p {\partial}
\def \BEA {\begin{eqnarray}}
\def \EEA {\end{eqnarray}}

\def \BC {\begin{cases}}
\def \EC {\end{cases}}

\usepackage[T2A]{fontenc}
\usepackage{hyphenat}%
\usepackage[english]{babel}

\begin{document}
\title
{ 
Large Fluctuations in Open Quantum Systems
}

\author{V.\,Yu. \!\!Mylnikov}
\email{vm@mail.ioffe.ru}
\affiliation{Ioffe Institute,
194021 St.~Petersburg, Russia}
\author{S.\,O.~Potashin}
\affiliation{Ioffe Institute,
194021 St.~Petersburg, Russia}
\author{A. Kamenev}
\affiliation{School of Physics and Astronomy, University of Minnesota, Minneapolis, MN 55455, USA  }
\affiliation{William I. Fine Theoretical Physics Institute, University of Minnesota, Minneapolis, MN 55455, USA }


\begin{abstract} 
We study statistics of atypical measurement outcomes in the steady states of driven open quantum systems. In equilibrium, the probability distribution over the phase space, as encoded in, e.g.,  the Wigner function, is analytic in the phase-space coordinates. We show that this property is generically lost in driven dissipative systems: their {\it large-deviation function} develops lines and surfaces across which its derivatives are discontinuous. As an illustrative example, we consider a parametrically driven Kerr oscillator coupled linearly and/or nonlinearly to a dissipative bath. Rare fluctuations in the amplitude and phase of the induced oscillations are governed by semiclassical instanton trajectories of the corresponding Keldysh-Lindblad action. We demonstrate that a given fluctuation can be realized through multiple distinct instanton trajectories. The competition between these trajectories leads to abrupt switching of the dominant instanton and, consequently, to non-analytic features in the large-deviation function.
\end{abstract}

\maketitle

\section{Introduction}

Studies of driven open quantum oscillators have attracted significant attention in the field of quantum information processing \cite{Proposal_Encode_Qubits, CatQubitBackground2, ExprQPDO2, CatQubitBackground3, Adiabatic_quantum_computing_nonlinear_network, Preparation_of_states_PDO, Driving_into_chaos_regime_validity,Exp_cancellation_tunneling,Exp_large_bit_flip,Exp_two_photon_dissipation}. These systems can provide novel platforms for sensing and  metrology \cite{Sensing_classical_1, Sensing_classical_2, Sensing_classical_3,Eichler2023, Quantum_metrology}. They can also serve as robust qubits with their non-equilibrium steady states encoding the logical basis  \cite{Proposal_Encode_Qubits, 2PhotonLossExp3, ExprQPDO2, Exp_cancellation_tunneling, CatQubitBackground2,CatQubitBackground3}. The quantum nature of these, essentially macroscopic, systems 
is ultimately responsible for the probabilistic nature of measurement  outcomes. Here we discuss generic features of far tails of the corresponding probability distributions, providing probabilities of rare atypical measurement outcomes.

To set the stage, we briefly recall known results for large-deviation functions in {\it classical} stochastic systems \cite{TOUCHETTE20091,Assaf_2017}. In thermal equilibrium, the probability of finding a system in a phase-space location  $(x, p)$  is given by the  Maxwell--Boltzmann distribution ${\cal P}(x,p)\propto e^{-E(x,p)/T}$, where $E(x,p)$ is the energy, e.g., $E(x, p)= p^{\,2}/2+U(x)$, and $T$ is the temperature of a bath, weakly coupled to the system.  The large deviation function, defined as:
\begin{equation}
        \label{eq:large-deviation-class}
    S(x,p)= \lim_{T\to 0} (-T)\ln {\cal P}(x,p), 
\end{equation}
coincides with the energy, $S(x,p)=E(x,p)$, and is therefore analytic throughout the phase space for smooth potentials $U(x)$. The situation changes qualitatively away from equilibrium. Stationary states, defined by the Fokker-Planck equation: 
$\partial_t {\cal P}=-\vec \nabla_{(x,p)}\cdot   {\cal {\vec J}}=0$,  can sustain a non-vanishing probability current $ {\cal {\vec J}}(x,p)$. This arises, for example, from adding phase-space dependent dissipative forces, which explicitly break time-reversal invariance and detailed balance.  
 In such cases, it has been show \cite{Graham1983,Dykman1994,Freidlin1998,Graham1985,Graham1983,Graham1984a,Graham1984b} that the large deviation function (\ref{eq:large-deviation-class})  can develop {\it non-analytic} lines in the phase space. Across such lines derivatives of (\ref{eq:large-deviation-class}) are discontinuous, even though all constituent forces are perfectly smooth functions of phase-space coordinates.

Within the large-deviation theory \cite{Freidlin1998,Dembo2009,Graham1984a,Graham1984b,Graham1985,Graham1983,Dykman1994} this behavior is understood as a manifestation of a fold catastrophe of the manifold of Lagrangian trajectories (instantons) associated with the Martin–Siggia–Rose action \cite{Martin1973}. Projection of such a folded manifold onto the physical phase space produces caustics which bound regions where multiple trajectories reach the same point $(x,p)$. In the weak noise limit, $T\to 0$ in Eq.~(\ref{eq:large-deviation-class}),  the probability ${\cal P}(x,p)$ is determined by a  trajectory with smallest action. The resulting non-analytic lines, located between the caustics, arise from abrupt switching between competing optimal trajectories.

Here we address rare events in open systems in the regime where {\it quantum} fluctuations dominate over thermal noise. In analogy with Eq.~(\ref{eq:large-deviation-class}), one defines quantum large-deviation function as
\begin{equation}
\label{eq:large-deviation-quant}
S(x,p)=\lim_{\hbar\to 0}(-\hbar)\ln {\cal P}(x,p),
\end{equation}
where $\hbar\to 0$ corresponds to the semiclassical limit of vanishing quantum fluctuations. In practice, this limit can be approached by tuning system parameters such as detuning or nonlinearity. Since quantum systems cannot be strictly localized at a point $(x,p)$ of its phase space, ${\cal P}(x,p)$ should be interpreted as a probability to fall into a phase-space region of area $\sim \hbar$ centered at $(x,p)$. To avoid ambiguities associated with this coarse-graining, we identify ${\cal P}(x,p)$ with the Wigner distribution \cite{Wigner1932}, $W_0(x,p)$, computed from the {\it stationary} reduced density matrix of the open system.  Although $W_0(x,p)$ is not in general positive definite, it remains non-negative in the examples considered here. We have further verified that the strictly positive Husimi distribution exhibits the same qualitative behavior.

Consider, e.g., a harmonic oscillator with frequency $\omega_c$
in its ground state. The corresponding Wigner distribution is $W_0(x,p) \propto e^{-2E(x,p)/\hbar \omega_c} $,  yielding the large deviation function (\ref{eq:large-deviation-quant}) as $S(x,p)=2E(x,p)/\omega_c$, where $E(x,p)$ is the classical oscillator energy. While this simple proportionality does not extend to anharmonic systems, their equilibrium (ground-state) large deviation functions remain analytic throughout the entire phase space \cite{Berry1977}.

Such analyticity does {\it not} extent to parametrically driven systems coupled to a bath. In an immediate vicinity of classically stable points, their stationary probability distribution closely resembles the thermal one with some effective temperature. This temperature is finite even if the bath temperature is strictly zero, and is of a purely quantum origin, i.e.  $\propto \hbar$.  This phenomenon is known as {\it quantum heating} \cite{Dykman2011,Dykman1988}. Yet, the global structure of the probability distribution is very different from that in thermal equilibrium. Most notably, the corresponding large deviation function (\ref{eq:large-deviation-quant}) exhibits non-analytic lines, where its derivative with respect to the phase-space coordinates are discontinuous.

The origin and quantitative structure of this phenomenon is the main focus of the present work. Although  non-analyticity of the large deviation function  is a common theme between  classical and quantum non-equilibrium systems, the details are rather different.  While in classical systems the non-analyticity requires an explicit breaking of the time-reversal symmetry, in quantum models it may coexist with (hidden) time-reversal symmetry \cite{Roberts2021}. Moreover, we found that the quantum one is not necessarily associated with the fold catastrophe of the Lagrangian manifold. Instead, their Lagrangian manifold forms a Riemann surface, glued of two copies of physical phase space. As a result, {\it any} phase-space point may be reached by two competing instanton trajectories, ending on different phase-space sheets.

 Another qualitative distinction is in the presence of  Stokes phenomenon in the quantum setup. It dictates that some instanton trajectories, although allowed by the equations of motion and boundary conditions, do not contribute to a probability of a rare event. The reason is that (functional) integration contours can't be deformed to pass through all stationary points (or rather trajectories) and some of them are left out. This translates into an intricate pattern of Stokes and anti-Stokes lines in the phase space, with the latter ones spelling the non-analyticity of the large deviation function.

This paper is structured as follows: Section \ref{sec:Model} introduces a model of a parametrically driven oscillator coupled to a bath and discusses its classical limit.   Section \ref{SectWigFun} is devoted to the Wigner function, where we discuss its exact form for models admitting hidden time-reversal symmetry \cite{Roberts2020,Roberts2021} and its WKB approximation. In Section \ref{Sec:instantons}, we formulate the real-time instanton approach within the Keldysh framework, compute the instanton trajectories, and link it to the Wigner distribution. The Stokes phenomenon is also discussed here and quantum phase-slip rate is evaluated. Finally,  Section \ref{sec:conclusions} presents our conclusions and discusses some open questions.  Technical details are delegated to three Appendices.

\section{The Model}
\label{sec:Model}

Consider a quantum oscillator (e.g., a mode of a resonant cavity) with the frequency $\omega_c$ and quartic anharmonicity. The oscillator is subjected to a parametric drive, i.e. modulation of its eigenfrequency, with frequency $2 \omega_{p}$. The {\it time-dependent} Hamiltonian of this system is given by:
\begin{equation}
\hat{H}(t)=
\frac{\hat{p}^2}{2}
+\frac{1}{2}\big[\omega_c^2+4\omega_c G\sin(2 \omega_{p} t)\big]\hat{x}^2+\frac{\omega_c^2}{3\hbar }\, U\hat{x}^4
,
\end{equation}
where a mass is absorbed into a rescaling of the momentum. To isolate slow amplitude dynamics, one employs the rotating wave approximation, \cite{Thompson2025}. To this end, one defines an annihilation operator as
\begin{equation} 
                    \label{eq:a}
\hat{a} = \sqrt{\frac{\omega_c}{2 \hbar}} \left(\hat{x}+i\frac{ \hat{p}}{\omega_c}\right) e^{i\omega_p t}. 
\end{equation}
This transformation allows one to rewrite the Hamiltonian in terms of variables rotating with the high frequency $\omega_p$. Neglecting rapidly oscillating contributions, yields an effective static Hamiltonian:
\begin{equation}\label{Hrwa} 
\hat{H}_0/\hbar = -\Delta\,\hat{a}^{\dagger}\hat{a} + \frac{iG}{2}\left(\hat{a}^{\dagger 2} - \hat{a}^{2}\right)+ \frac{U}{2}\,\hat{a}^{\dagger 2} \hat{a}^{2}, \end{equation}
where the bosonic operators $\hat{a}$ and $\hat{a}^{\dagger}$ satisfy the canonical commutation relation,  $[\hat{a},\hat{a}^{\dagger}] = 1$,   $\Delta = \omega_{p} - \omega_{c}-U$ is a detuning between the (half of) the pump and oscillator frequencies, with the nonlinear frequency shift taken into account, $G$ is a coherent two-photon drive amplitude, and $U$ is Kerr nonlinearity. 

To avoid heating of a driven non-linear  system all the way up to $T=\infty$, one must allow for energy exchange with a bath. 
Given the high operating frequency, $\omega_p$, one may  assume that the density of states of a generic bath is not singular in the vicinity of $\omega_p$. This  justifies \cite{Kamenev} using the Markovian approximation for the system--bath interactions, which results in the Lindbladian form of the evolution equation for the {\it reduced} density matrix of the system 
\begin{equation}\label{Lindblad}
\begin{gathered}
\partial_{t}\hat{\rho}
= -i[\hat{H}_0,\hat{\rho}]/\hbar
+ \sum\limits_{\hat L} \kappa_{\hat L}\,\mathcal{D}_{\hat L}[\hat{\rho}] ,
\end{gathered}
\end{equation}
where $\mathcal{D}_{\hat{L}}[\hat{\rho}] = \hat{L}\hat{\rho}\hat{L}^{\dagger}
- \{\hat{L}^{\dagger}\hat{L},\hat{\rho}\}_+/2$ are dissipative superoperators, specified by jump operators $\hat L$, and $\kappa_{\hat L}>0$ are the corresponding dissipation rates. 
The set of jump operators and corresponding rates considered here is as follows: 
\begin{equation}
                \label{eq:jump-operators}
\begin{aligned}
&\hat L =\hat a; \qquad \quad \kappa_{\hat a} =2\kappa;\\
&\hat L =\hat a^2;\quad \,\,\,\quad \kappa_{\hat a^2} =\eta;\\
&\hat L =\hat a^\dagger \hat a;\qquad \kappa_{\hat a^\dagger\hat a} =\kappa_\phi.
\end{aligned}
\end{equation}
The first two represent one- and two-photon loss, respectively, while the third one -- loss of coherence (dephasing) of the cavity oscillations. One can add one- and two-photon gain processes, $\hat L =\hat a^\dagger$ and $\hat L =\hat a^{\dagger 2}$. Those are associated with a finite temperature, $T$, of the bath, so, e.g., $\kappa_{\hat a^\dagger}/\kappa_{\hat a} = e^{-\hbar \omega_c/T}$. Since we are mostly interested in the deep quantum limit $T\ll \hbar \omega_c$, we do not include photon gain processes.

It is worth noticing that as far as scaling with $\hbar$ is concerned,  $G,\kappa,\Delta\sim \hbar^0$, while $U,\eta,\kappa_\phi\sim \hbar$. 
To understand the {\it classical} limit of the system's dynamics, one writes the equation of motion for the average coherent field amplitude $\alpha(t)=\langle  \hat{a}\rangle =\mathrm{Tr}\{\hat{a}\,\hat\rho(t)\}$: 
\begin{equation}
 \p_t \alpha=i( \Delta+i\kappa+i\kappa_\phi/2) \alpha+  G{\alpha}^*-(\eta+i U) \langle \hat{a}^{\dagger} \hat{a}^2\rangle, 
\end{equation}
Using the fact that $\hat{a},\hat{a}^\dagger\sim\hbar^{-1/2}$ (cf. \eqref{eq:a}) one can replace the expectation values of the operator products by the products of their expectation values $\langle \hat{a}^{\dagger} \hat{a}^2\rangle\approx{\alpha}^*\alpha^2$. 
In the limit $\hbar\to 0$, one thus obtains the 
classical equation of motion \cite{Bartolo2016}:
 \begin{equation}\label{alpha}
 \p_t \alpha=i\tilde{\Delta}^*\alpha+  G{\alpha}^*-\tilde{\eta}^* {\alpha}^*\alpha ^2 ,
\end{equation}
where   $\tilde{\Delta}=\Delta-i\kappa$ is complex detuning and \mbox{$\tilde{\eta}=\eta-iU$} -- complex nonlinearity. Note that the contribution associated with the dephasing, $\sim\kappa_{\phi}$, does not enter the classical equation  \eqref{alpha}, because it scales proportional to the Planck constant. 

Equation~\eqref{alpha} admits several fixed points \cite{Meaney2014}. We restrict our analysis to the case $\rm{Im}(\tilde{\Delta}/\tilde{\eta})<0$, which yields one or three fixed points. In the weak drive  regime $G<|\tilde{\Delta}|$  there is only one stable point $\alpha=0$. This corresponds to the absence of stable  parametric oscillations. 
In the opposite limit $G>|\tilde{\Delta}|$ there are three fixed points: an unstable saddle point $\alpha=0$ and two stable points $\alpha=\pm\alpha_0$, where $\alpha_0=\sqrt{n_0}\,e^{i \theta_0}$: 
\begin{equation}\label{averagen}
 n_0=\sqrt{\frac{G^2}{|\tilde{\eta}|^2}-\rm{Re}\left(\frac{\tilde{\Delta}}{\tilde{\eta}}\right)^2 }+\rm{Im}\left(\frac{\tilde{\Delta}}{\tilde{\eta}}\right), 
 \end{equation}
\begin{equation}
\label{eq:theta}
 e^{i\theta_0}=\sqrt{\frac{n_0+i\tilde{\Delta}/\tilde{\eta}}{G/\tilde{\eta}}}.
\end{equation}
This describes classically (bi)stable parametrically induced oscillations with amplitude $\sqrt{n_0}$ and phases $\theta_0$ and $\theta_0-\pi$ relative to the drive phase. The transition from a single fixed point to the three fixed points is a bifurcation, also known as a dissipative phase transition \cite{Bartolo2016,Mylnikov2022,Downing2023,Mylnikov2025}.   

In the following sections, we investigate the quantum dynamics of the system, focusing on its long-time stationary state.  To this end, we first employ the Wigner function, which allows one to find exact solutions in some particular cases. We then show how these results follow from the real-time instanton formalism, which is applicable in more general situations.   

\section{Wigner function }
\label{SectWigFun}

To analyze and visualize the reduced density matrix, it is convenient to construct a phase-space distribution function \cite{Walls,Scully,Cahill1969}. The  most common ones are:  Husimi Q-function \cite{Husimi1940}, P-function \cite{Glauber1963,Walls,Drummond1980II,Bartolo2016}, and Wigner function.   
Here we focus on the Wigner distribution function \cite{Wigner1932}, because it provides  an intuitive phase-space description of quantum states, and because it naturally emerges in the coherent state path integral description of the Lindbladian evolution \cite{Kamenev}. Unlike the complex P-representation, the Wigner function is manifestly real-valued. It is worth mentioning that (unlike for typical events) the large-deviations parts of all these representations are practically indistinguishable.  

The Wigner function is expressed through the density matrix  in the following way \cite{Scully}: 
\begin{equation}\label{WignerExact}
W(\alpha,{\alpha}^*;t)=\mathcal{N}\,\text{Tr}\big\{ \hat{D}(\alpha)\,e^{i \pi \hat{a}^\dagger \hat{a} }\, \hat{\rho}(t)\,\hat{D}^\dagger(\alpha)\big\},
\end{equation}
where $\hat{D}(\alpha)=\exp\{\alpha \hat{a}^\dagger-{\alpha}^* \hat{a}\}$ is a phase-space displacement operator and $\mathcal{N}$ is a normalization constant. Starting from the Lindblad equation \eqref{Lindblad}, a corresponding evolution equation for the Wigner function can be obtained using the operator correspondence rules  \cite{Scully}. Those map  operators acting on the density matrix  onto  differential operators acting on the phase-space variables $\alpha$ and $\alpha^*$ according to: 
\begin{equation}
\begin{aligned}
&\hat{a}\hat{\rho} \rightarrow
\left(\alpha+\frac{1}{2}\frac{\partial}{\partial \alpha^*}\right)W;\quad
\hat{a}^{\dagger}\hat{\rho} \rightarrow
\left(\alpha^*-\frac{1}{2}\frac{\partial}{\partial \alpha}\right)W;\\
&\hat{\rho}\hat{a} \rightarrow
\left(\alpha-\frac{1}{2}\frac{\partial}{\partial \alpha^*}\right)W;\quad
\hat{\rho}\hat{a}^{\dagger} \rightarrow
\left(\alpha^*+\frac{1}{2}\frac{\partial}{\partial \alpha}\right)W .
\end{aligned}
\end{equation}
Applying these rules to the Hamiltonian and dissipative parts of Eq.~\eqref{Lindblad}, one obtains the evolution equation for the Wigner function:
\begin{equation}\label{WignerEq}
\begin{gathered}
 \frac{\partial W}{\partial t} = 
\left\{-\frac{\p}{\p \alpha^*}\left(- i\tilde\Delta \alpha^* + G\alpha- \tilde\eta\alpha\alpha^{*2}\right) 
+\right. \\\left.\frac{1}{2}\frac{\p}{\p \alpha^*}\left[\left((2\tilde\eta+\kappa_{\phi})\alpha\alpha^*+i\tilde{\Delta}\right)\frac{\p}{\p \alpha}-\kappa_\phi\alpha^{*2}\frac{\p}{\p \alpha^*}\right]+\right. \\\left.
\frac{\tilde{\eta}}{4}\frac{\p}{\p \alpha^*}\alpha  \frac{\p^2}{\p \alpha^2}
+ c.c.\right\} W.
\end{gathered}
\end{equation}
The first term on the right side of Eq. \eqref{WignerEq} represents  the classical drift in phase space under the action of external forces, cf. Eq.~\eqref{alpha}, while the second term accounts for diffusion. The third term, which contains a third-order derivative, does not have a classical analog in the Fokker-Planck equation. This term is a fundamental signature of non-classicality and significantly complicates the analysis of both the steady-state properties and dynamics of the system. A common approach to simplifying Eq. \eqref{WignerEq} is the truncated Wigner approximation \cite{Walls, Kinsler1991}, which involves neglecting the third-order derivatives in Eq. \eqref{WignerEq}. However, it overestimates the decoherence rate \cite{Kinsler1991} and thus cannot provide a correct description of the complete quantum problem. This study deals with these obstacles by considering the exact Wigner function.

\begin{figure*}
\includegraphics[width=0.95\textwidth]{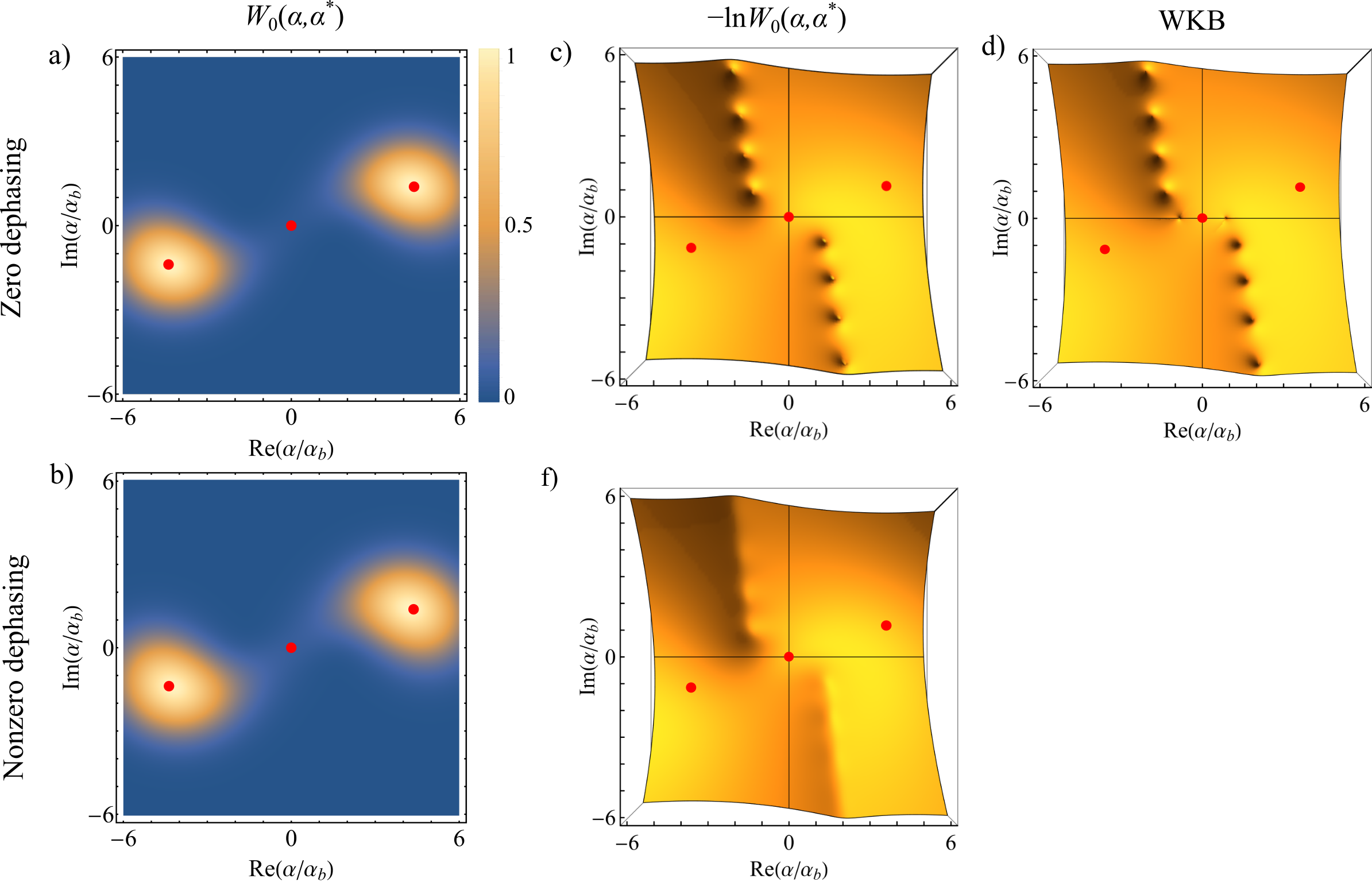}
\caption{(a,b) Stationary Wigner function $W_0(\alpha,\alpha^*)$ and effective potential $-\ln W_0(\alpha,\alpha^*)$ (c,d) vs. phase-space coordinates $\rm{Re}(\alpha)$ and $\rm{Im}(\alpha)$. The first row presents the case of zero dephasing ($\kappa_\phi=0$), where the exact (a,c) and WKB (d) solutions present (see Eqs. \eqref{Decomp}-\eqref{PsiExact} and Appendix \ref{AppWKB}). The second row shows the nonzero dephasing case ($\kappa_\phi=0.2$), which does not admit the holomorphic decomposition \eqref{Decomp} and is calculated numerically. The system parameters are set to $G=5, \Delta=2, \kappa=0.4, \eta=1,U=0.1$. The red dots denote locations of the classical fixed point \eqref{averagen}-\eqref{eq:theta}.
}
\label{FigLogW}
\end{figure*}

First, we seek some guidance from the  Wigner function equation \eqref{WignerEq}. An exact solution can be obtained in the special limit of $\kappa_\phi=0$ (and the absence of finite $T$ photon-gaining jump operators). As was first realized by Roberts, Lingenfelter, and Clerk \cite{Roberts2021}, such a model admits a hidden time-reversal symmetry (HTRS), which considerably simplifies its analytic treatment. Its exact stationary Wigner function was found in a somewhat indirect way using the P-representation \cite{Bartolo2016} and the quantum absorber method \cite{Roberts2020}. 
Here we show that the stationary Wigner function may be directly deduced from Eq. \eqref{WignerEq}, by looking for its solution in the  factorized holomorphic form:
\begin{equation}\label{Decomp}
W_0(\alpha,\alpha^*) =\mathcal{N}_0\,  e^{-2\alpha^*\alpha}\, \Psi(\alpha) {\Psi}^*(\alpha^*),
\end{equation}
where $\Psi(\alpha)$ and ${\Psi}^*(\alpha^*)$ are holomorphic functions and the normalization coefficient is given by $\mathcal{N}_0^{-1}=e^{-2\alpha_0^*\alpha_0} \Psi(\alpha_0) {\Psi}^*(\alpha_0^*)$.  
Notice that the holomorphic factorization (\ref{Decomp}) does not extend to time-dependent solutions of Eq.~(\ref{WignerEq}).   Substituting ansatz ~\eqref{Decomp} into Eq.~\eqref{WignerEq} with $\kappa_\phi=0$, one obtains a closed equation for  $\Psi(\alpha)$: 
\begin{equation}\label{difeq1}
\begin{gathered}
\alpha \tilde{\eta} \,\Psi''(\alpha) + 2 i \tilde{\Delta} \Psi'(\alpha) - 4 \alpha G \Psi(\alpha) = 0,
\end{gathered}
\end{equation}
and its conjugate one for ${\Psi}^*(\alpha^*)$. A very similar equation appears in Ref.~\cite{Roberts2020} as the Segal-Bargmann representations of the pure-state wave function in the coherent quantum-absorber method.   Solution of Eq.~\eqref{difeq1}, {\it regular} at $\alpha=0$,  is given by 
\begin{equation}\label{PsiExact}
\Psi(\alpha)=\ _0F_1\left(\frac{1}{2}+\frac{i \tilde\Delta
   }{\tilde\eta };\,\frac{G}{\tilde\eta
   } \,\alpha^2 
  \right),
\end{equation}
where  $_0F_1(a;z)$ is the confluent hypergeometric function. The corresponding Wigner function \eqref{Decomp} agrees with that of Refs.~\cite{Bartolo2016,Roberts2020}. 

Figure \ref{FigLogW}a shows the contour plot of the corresponding stationary Wigner function. As expected, the probability is heavily concentrated around two oscillatory modes with $\alpha=\pm \alpha_0$. The faint tails towards $\alpha =0$ indicate the small probability that the system may be found around that point and thus undergo a phase-slip between the two quasi-stable oscillatory states. Such phase-slips are the main source of qubit decoherence \cite{Marthaler2007,Dykman1994,Dykman1998,Dykman2007,Dykman2012,Su2025,Frattini2024,Boness2025,Putterman2025,Dubovitskii2024,Dubovitskii2025,Chamberland2022,Kinsler1991,Sun2019,Lee2025,Thompson2022,Sepulcre2026,Carde2025,Thompson2026}. We address them in Section \ref{sec:phase-slips}. 

In analogy with the Boltzmann distribution, the logarithm of the stationary Wigner function, Fig.\ref{FigLogW}c, may be called an effective potential:
\begin{equation}
\begin{gathered}\label{EfPotbase}
S(\alpha,\alpha^*)=-\ln W_0(\alpha,\alpha^*).
\end{gathered}
\end{equation}
Predictably, it exhibits the two degenerate minima at $\alpha=\pm\alpha_0$, and the saddle point  at $\alpha=0$. However, the most noticeable feature of it is the two sharp ``mountain range'' lines, stretching from the two opposite points 
\mbox{$\alpha =\pm \alpha_b$} towards infinity, where $\alpha_b=\tilde{\Delta}/(2\sqrt{G\tilde{\eta}})$. As we show below, the quantum large deviation function (\ref{eq:large-deviation-quant}), where the probability ${\cal P}$ is associated with the Wigner function, is non-analytic along these ``mountain range'' lines with discontinuous first (and higher) derivatives.  

To understand this phenomenon, let us look at the asymptotic behavior of function $\Psi(\alpha)$. The limit \mbox{$\hbar\to 0$} corresponds to  a small nonlinear parameter $|\tilde\eta|\ll G, |\tilde\Delta|$, making the coefficient in front of the highest derivative in Eq.~(\ref{difeq1}) small. Thus, one can apply WKB methods to obtain asymptotic estimates for the stationary Wigner function. This procedure (see Appendices \ref{AppWKB} and \ref{AppComPlWKB} for details) leads to the following expression: 
\begin{equation}
\begin{gathered}
                    \label{eq:asymptotic}
\Psi(\alpha) \propto 
\Psi_{-}(\alpha)+\Psi_{+}(\alpha),
\end{gathered}
\end{equation}
where the two asymptotic components are given by 
\begin{equation}\label{eq:Psi-pm}
\Psi_\pm(\alpha) = A_{\pm}(\alpha/\alpha_b)\exp \left\{i\delta\,\phi_{0}^{(\pm)}(\alpha/\alpha_b)\right\},
\end{equation}
where $\delta=\tilde \Delta/\tilde \eta \propto \hbar^{-1}$ is the large WKB parameter. The two WKB actions are given by 
\begin{equation}\label{Eq:phiMapp}
    \phi_{0}^{(-)}(z)= \sqrt{1- z ^2}
- \ln  \left(1+\sqrt{1- z^2}\right);
\end{equation}
\begin{equation}\label{Eq:phiPapp}
\hskip -1.6cm \phi_{0}^{(+)}(z)= -\phi_{0}^{(-)}(z)-
\ln(z^2),
\end{equation}
where $z=\alpha/\alpha_b$. 
The slowly changing pre-exponential factors $A_{\pm}(z)$ are derived in Appendix \ref{AppWKB}.

Notice that, while $\Psi_{-}(\alpha)$ is regular in $\alpha=0$, the second asymptotic component, $\Psi_{+}(\alpha)$, is singular due to the presence of $\ln(z^2)$ in \eqref{Eq:phiPapp}. This indicates that, because of the Stokes phenomenon, there is a region containing $z=0$, where  $A_+(z)\equiv 0$ and thus $\Psi(\alpha) \propto 
\Psi_{-}(\alpha)$, see Section \ref{sec:Stokes} for more details. The cut of $\ln(z^2)$ in \eqref{Eq:phiPapp} should be understood as being chosen inside this Stokes region. Therefore, the exact location of the cut does not influence $\Psi(\alpha)$, since $\Psi_{+}(\alpha)=0$ in the Stokes region.  Fig. \ref{FigLogW}d shows the minus logarithm of WKB approximation for the Wigner function. It is almost indistinguishable from the  exact result, Fig.~\ref{FigLogW}c.

Now one can understand the origin of the ``mountain ranges'' as a result of the competition (taking place outside of the Stokes region) between the two asymptotic components $\Psi_\pm(\alpha)$. Switching between two WKB solutions occurs when the real parts of the exponents in Eq. \eqref{eq:Psi-pm} coincide, i.e., when 
\begin{equation}
    \label{eq:anti-Stokes}
    \mathrm{Im}\left\{ \delta\, \phi_{0}^{(+)}(\alpha/\alpha_b)\right\} = \mathrm{Im}\left\{\delta \, \phi_{0}^{(-)}(\alpha/\alpha_b)\right\}.
\end{equation}
This condition defines the anti-Stokes lines on the $\alpha$-plane, which faithfully follow the "mountain range" structures observed in Fig. \ref{FigLogW}c,d. The large deviation function (\ref{eq:large-deviation-quant}) exhibits discontinuous derivative along the lines specified by Eq.~(\ref{eq:anti-Stokes}). Indeed, in the limit $\hbar\to 0$, i.e. $|\delta|\to \infty$, only one asymptotic component with the smallest (negative) real part of the exponent dominates the observed probability. Such leading component switches abruptly at the lines (\ref{eq:anti-Stokes}). Thus, while the large deviation function is continuous, all its derivatives exhibit discontinuous jumps.        
 
The sharp peaks along these switching lines can be understood as interference effects between the two solutions in Eq. \eqref{eq:asymptotic}.  Although the $\Psi_\pm$ asymptotics share the same amplitude along the mountain lines \eqref{eq:anti-Stokes}, they exhibit different phases, thus the Wigner function reaches zero in isolated points along \eqref{eq:anti-Stokes} where $A_+\mathrm{e}^{i\,\mathrm{Re}\,\{\delta\,\phi_{0}^{(+))}\}} + A_-\mathrm{e}^{i\,\mathrm{Re}\{\delta\,\phi_{0}^{(-)}\}}=0$, leading to a divergence of its (negative) logarithm. 

One may wonder if the presence of non-analytic lines in the quantum large deviation function (\ref{eq:large-deviation-quant}) is a feature of the models with hidden time-reversal symmetry \cite{Roberts2021}, which admit the holomorphic factorization (\ref{Decomp}) of the stationary Wigner distribution. To address this, we numerically evaluated the stationary Wigner function for a model with $\kappa_\phi > 0$, which does not admit factorization (\ref{Decomp}). The result is presented in Fig.~\ref{FigLogW}b,f. One can clearly see that the non-analytic lines are still present. On the other hand, the sharp peaks along these lines are not there.  

In the next Section we interpret the non-analytic lines in the large deviation function from the real-time instanton perspective. This consideration shows that they are generic features of open driven quantum systems. Their origin is traced to the presence of multiple semiclassical trajectories that reach a given observation point $(\alpha,\alpha^*)$. The non-analytic lines trace the regions of the phase space where the actions along competing trajectories equalize. This leads to an abrupt switch of the most probable trajectory, resulting in discontinuous derivatives of the large deviation function (\ref{eq:large-deviation-quant}).    

\section{Real-time Instantons}
\label{Sec:instantons} 

One may describe the Lindbladian evolution \eqref{Lindblad} as the Keldysh field theory in the coherent state representation \cite{Kamenev,Sieberer2016}. The {\it probability} of finding the system in a certain region of its phase space centered at $\alpha$  can be found as the path integral: 
\begin{equation}
        \label{eq:funct-integral}
W_0(\alpha,\bar\alpha)=\iiiint\limits^{\alpha,\bar\alpha}\! \mathcal{D}[\alpha_{cl},\bar\alpha_{cl},\alpha_{q},\bar\alpha_{q}]\, \, e^{- S[\alpha_{cl},\bar\alpha_{cl},\alpha_{q},\bar\alpha_{q}]}, 
\end{equation}
where $\mathcal{D}[\alpha_{cl},\bar\alpha_{cl},\alpha_{q},\bar\alpha_{q}]$ 
is the flat functional measure and $S$ is  {\it local-in-time} Keldysh action: 
\begin{equation}\label{action}
 S =
 \!\int\! dt
 \big[ 
 \bar{\alpha}_{q}\p_t \alpha_{cl}
 -\alpha_{q}\p_t \bar{\alpha}_{cl}- \mathcal{L}(\alpha_{cl},\bar\alpha_{cl},\alpha_{q},\bar\alpha_{q})
 \big].
\end{equation}
Here $\alpha_{cl}(t)$ and $\alpha_{q}(t)$ are the classical and quantum components of the complex bosonic field, and the local-in-time function $\mathcal{L}(\alpha_{cl},\bar\alpha_{cl},\alpha_{q},\bar\alpha_{q})$ encodes both the Hamiltonian and dissipative parts of Eq.~(\ref{Lindblad}), according to the standard rules \cite{Sieberer2016,Kamenev}, see also Appendix \ref{AppKpi}.   The classical field trajectory,  $\alpha_{cl}(t)$, is the physical degree of freedom and should obey the boundary condition of reaching the observation point $\alpha_{cl}(t_f)=\alpha$, $\bar\alpha_{cl}(t_f)=\alpha^*$ at the final time $t_f$, with $t_f\to \infty$ -- if one is interested in the stationary state. 

The quantum field, $\alpha_{q}(t)$, encodes fluctuations, both quantum and thermal. One can sort the action by powers of the quantum field. There are no terms with zeroth power of $\alpha_{q}$.  The contributions that are linear in quantum fields encode classical equations of motion. The terms quadratic in the quantum fields may be treated as stochastic Langevin noise in the classical equations, resulting in the phase-space diffusion. Importantly, in  quantum problems with nonlinear interactions, the action \eqref{action} also includes contributions with higher powers of the quantum field, which have no analog in classical stochastic systems with a Gaussian noise and must be kept. For example, Kerr nonlinearity or two-photon dissipation produces terms that are third order in the quantum field \cite{Thompson2022,Carde2025}. This significantly complicates the calculation of the path integral. 

Here we  focus on the semiclassical limit, \mbox{$\hbar\rightarrow 0$}, where  the functional integral is dominated by stationary–instanton–field configurations  \cite{Thompson2022,Kamenev}. To implement this program, one looks for solutions of the saddle-point equations with proper boundary conditions. The corresponding equations are obtained by the variation of the Keldysh action with respect to quantum/classical fields and take the Hamiltonian form:
\begin{equation}\label{EqofM}
\begin{split}
       &d\alpha_{cl}/dt=\p \mathcal{L}/\p\bar{\alpha}_q;\qquad  d\bar{\alpha}_{q}/dt=-\p\mathcal{L}/\p \alpha_{cl}, 
       \\ 
       &d\bar{\alpha}_{cl}/dt=-\p\mathcal{L}/\p \alpha_q; \quad\,
       d\alpha_q/dt=\p\mathcal{L}/\p\bar{\alpha}_{cl}, 
\end{split}
\end{equation}
which conserves the $\mathcal{L}$-function ($d \mathcal{L}/dt=0$), playing the role of "energy" or an {\it effective} Hamiltonian. 

Nominally, $\bar{\alpha}_{cl/q}$ are the complex conjugates of $\alpha_{cl/q}$, respectively. However, the saddle-point equations \eqref{EqofM} do not respect this structure (see Appendix \ref{AppKpi} for details). Consequently, to access the saddle point, one needs to deform the integration contours in the {\it complex} planes of $\mathrm{Re}\,\alpha_{cl}$, $\mathrm{Im}\,  {\alpha}_{cl}$, $\mathrm{Re}\,\alpha_q$, $\mathrm{Im}\,  {\alpha}_q$, treating all of them as four independent complex variables. 
This effectively doubles the dimensionality of a space available for quasi-classical trajectories, making it 8-dimensional. However, as noticed in~\cite{Kamenev}, there exists a 4-dimensional invariant subspace, which hosts all trajectories of interest. It is parametrized by two complex fields $\alpha(t)$ and $\chi(t)$, defined as 
\begin{equation}
    \label{eq:chi-variable}
\alpha=\alpha_{cl},  \quad \bar\alpha= \bar\alpha_{cl};  \qquad \chi = \bar{\alpha}_q, \quad \bar{\chi}=-\alpha_q.
\end{equation}
As shown in Appendix \ref{AppKpi} for an arbitrary Lindbladian, the real-time equations of motions \eqref{EqofM} can always be satisfied with $\bar\alpha(t) =\alpha^*(t)$, and $\bar\chi(t)=\chi^*(t)$, bringing the effective dimensionality back to four real or two complex dimensions.   

For the parametric oscillator (\ref{Lindblad}) the corresponding {\it real } Lindbladian density, expressed in the new variables, takes the form  
\begin{equation}\label{Li1}
\begin{aligned}
    \mathcal{L}(\alpha,\bar\alpha,\chi,\bar\chi)&=\bar{\chi}\big[G\alpha- i\tilde{\Delta}(\bar{\alpha}-\tfrac{1}{2}\chi)-\tilde{\eta}\alpha(\bar{\alpha}-\tfrac{1}{2}\chi)^2
   \\&-\tfrac{1}{2}\kappa_{\phi}\bar{\alpha}  (\bar{\alpha}\bar{\chi}-\alpha  \chi )\big]+c.c.,
\end{aligned}
\end{equation}
where $c.c.$ stays for the complex conjugation. The corresponding equations of motions are 
\begin{equation}\label{EqofM1}
\begin{split}
       &d\alpha/dt=\p \mathcal{L}/\p\chi;\quad
        d\chi/dt=-\p\mathcal{L}/\p\alpha,
\end{split}
\end{equation}
and their complex conjugated. In the classical limit ($\chi = 0$), the saddle-point equations \eqref{EqofM1} reduce to the classical equation of motion \eqref{alpha}. As discussed in Section \ref{sec:Model}, the system always has at least one stable point in the classical plane $\chi = 0$. However, these classically stable points necessarily become metastable when the whole $(\alpha,\chi)$ complex phase space is taken into account. Physically, this means that quantum fluctuations can push the system out of the classical subspace, into a region where the quantum field is finite ($\chi \neq 0$). 

We aim to understand the steady state of the Lindbladian evolution approached at $t_f\to \infty$.  In the semiclassical approximation it is described by trajectories which take an arbitrarily long time to reach the observation point $\alpha$. All such trajectories must depart from  one of the classical stable points ($\alpha=\pm \alpha_0,\text{ or }\alpha=0, \chi=0$). Equations of motions \eqref{EqofM1}, linearized around one such point, admit four eigen-directions: two convergent ones belong to the classical plane $\chi=0$, while the remaining two diverge from the fixed point and have components along all four $(\alpha,\bar\alpha,\chi,\bar\chi)$ directions.  The departing trajectories start as linear combinations of the latter two eigen-directions and thus form a two-dimensional manifold. Since the effective Hamiltonian, $\mathcal{L}$, is conserved, the trajectory manifold is at a fixed zero ``energy'':
\begin{equation}\label{Lzero}
  \mathcal{L}[\alpha,\bar{\alpha},\chi,\bar{\chi}] = 0, 
\end{equation}
because the starting point belongs to the classical plane $\chi=\bar\chi=0$. Notice that  (\ref{Lzero}) does not fully specify the trajectory manifold. Indeed, since $\mathcal{L}$ is real, (\ref{Lzero}) only defines a three-dimensional subspace to which the two-dimensional trajectory manifold belongs.   

We are interested in a relative weight of finding  the system in a $\hbar$ vicinity of some point $\alpha$ of its phase space. To this end we need to find all trajectories which satisfy Eqs.~\eqref{EqofM1}, depart from  classically stationary points, and arrive (in arbitrarily long time) at the observation point $\alpha$. As explained above, all such trajectories belong to the two-dimensional manifold, embedded in the subspace specified by Eq.~\eqref{Lzero}.  The corresponding Wigner function in the semiclassical approximation is given by: 
\begin{equation}\label{WigAct}
  W_0(\alpha,\bar{\alpha}) = \sum_k B_k \exp \big\{ -S_k(\alpha,\bar{\alpha})\big\},
\end{equation}
where the sum runs over trajectories satisfying equations of motion \eqref{EqofM1} with the boundary conditions, specified above, and allowed by Stokes conditions. The latter ensure that the integration contour may be deformed to pass through the corresponding saddle point, see Section \ref{sec:Stokes}. The action along such a trajectory $(\alpha_k(t),\chi_k(t))$ is 
\begin{equation}\label{effpot}
S_k(\alpha,\bar{\alpha}) =\int dt\ \big[ \chi_k(t) \partial_t {\alpha}_k(t) +\bar{\chi}_k(t) \partial_t{\bar{\alpha}}_k(t) \big],
\end{equation}
where we used the condition $\mathcal{L}=0$ . Finally, $B_k$ are pre-factors coming from Gaussian integrations around the stationary trajectories. Explicitly finding the proper trajectories, in general, may be achieved only numerically. However, significant lessons may be learned from the case with hidden time-reversal symmetry (HTRS) \cite{Roberts2021},  where an analytical treatment is available.

\subsection{Hidden time-reversal symmetry case }

\begin{figure*}
\includegraphics[width=1.00\textwidth]{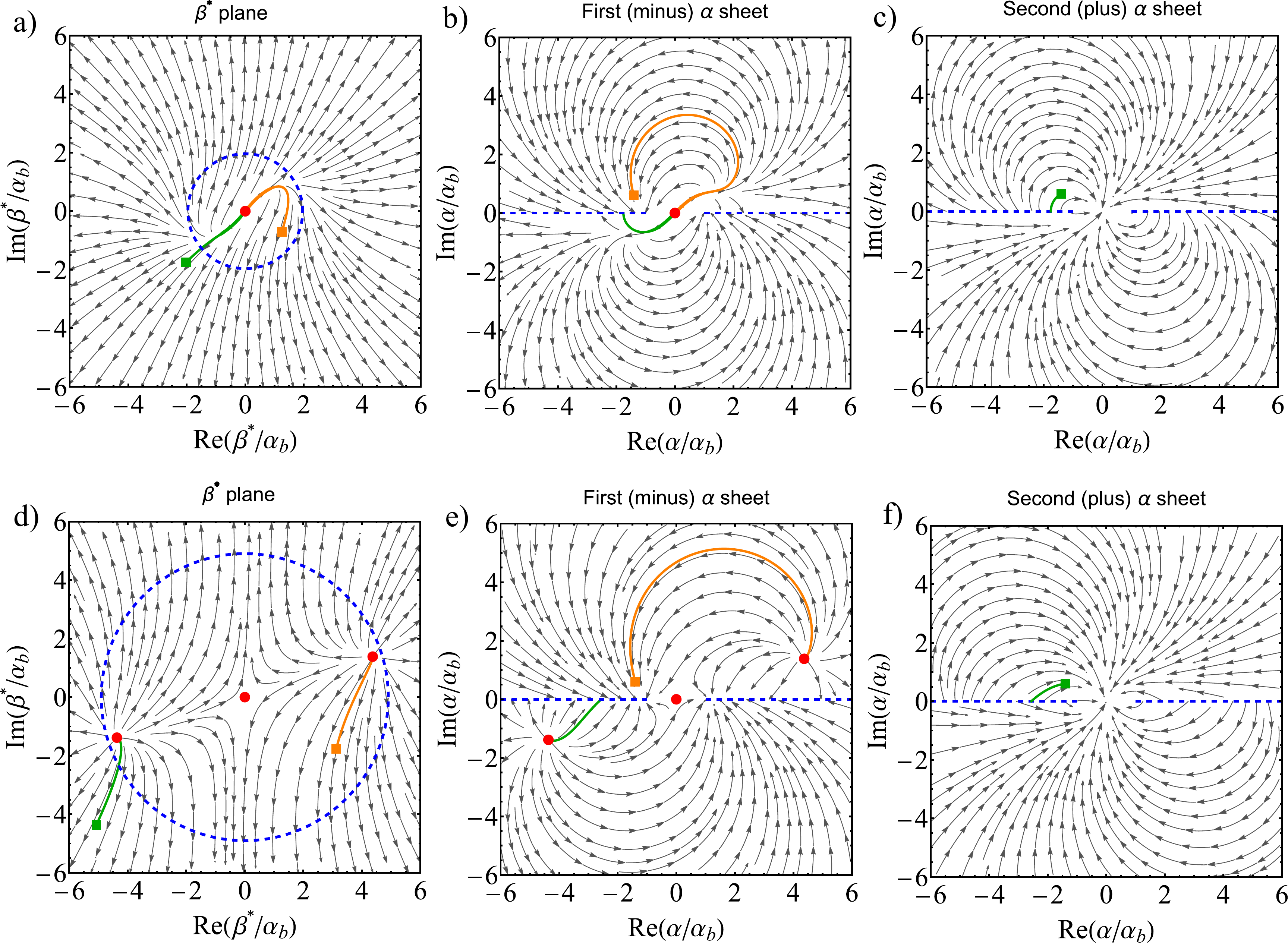}
\caption{The real-time instanton trajectories, Eq.~\ref{EqAlpha1}, spanning the fluctuation Riemann surface, ${\cal A}(\alpha,\beta)=0$. The instanton paths start at classical stable points (red dots). (a-c) The subcritical drive, $G<|\tilde\Delta|$; (d-e) -- the bistable regime,  $G>|\tilde\Delta|$.  (a), (d) Trajectories projections on the $\beta$-plane;  for a given end-point $\beta$ there is a unique trajectory, leading to it.  (b,c) and (e,f) Trajectories projections on the two sheets of the  $\alpha$-plane. There are two trajectories leading to each end-point $\alpha$: they both start on the minus-sheet (b), (e), one (orange) stays on this sheet, while the second (green) hits the cut (blue dashed line) and emerges on the second (plus) sheet (c), (f).    The system parameters are set to $\Delta=2, \kappa=0.4, \eta=1, U=0.1, \kappa_\phi=0$; (a-c) $G=2$; (d-f) $G=5$.}
\label{FigPhasePortriet}
\end{figure*}

In the case of zero temperature (i.e. no $a^\dagger$ or $a^{\dagger 2}$ jump operators) and no dephasing, $\kappa_\phi=0$, the Lindbladian density \eqref{Li1} acquires a characteristic form: 
\begin{equation}\label{Li2}
\begin{split}
&\mathcal{L} = 
\bar\chi\, \mathcal{A}(\alpha,\bar\alpha-\tfrac{1}{2}\chi) +
\chi\, \mathcal{A}^*(\bar\alpha,\alpha-\tfrac{1}{2}\bar\chi), 
\end{split}
\end{equation}
where we introduced a holomorphic polynomial: 
\begin{equation}
\begin{split}
                    \label{Eq:A}
\mathcal{A}(\alpha,\beta)=G \alpha -i\tilde{\Delta} \beta   -\tilde{\eta}\alpha \beta^2,
\end{split}
\end{equation}
of two complex variables $\alpha$ and $\beta=\bar\alpha-\tfrac{1}{2}\chi$. As a result, the $\mathcal{L}=0$ condition may be satisfied (in addition to the classical plane $\chi=0$, i.e. $\beta=\bar\alpha$) by 
\begin{equation}
\begin{split}
                    \label{Eq:Azero}
\mathcal{A}(\alpha,\beta)=0, 
\end{split}
\end{equation}
which specifies a {\it Riemann surface} (with genus zero). Let us verify that this Riemann surface is indeed an invariant manifold of equations of motion \eqref{EqofM1}.  To this end, we perform a canonical change of variables $(\alpha,\chi)\to(\alpha,2\beta)$ and find 
\begin{equation}\label{AtimeEvol}
\p_t\mathcal{A}=\dot{\alpha}\,\p_{\alpha}\mathcal{A}+\dot{\beta}\,\p_{\beta}\mathcal{A},
\end{equation}
where 
\begin{equation}\label{EqAlpha1}
    \begin{gathered}
    \dot{\alpha}=-\p_\beta\mathcal{L}/2={\mathcal{A}}^*-(\alpha-\bar{\beta})\p_{\beta}\mathcal{A};\\
\!\!\!\!\!\!\!\!\dot{\beta}=\p_\alpha\mathcal{L}/2=\mathcal{A}+(\alpha-\bar{\beta})\p_\alpha\mathcal{A}.
     \end{gathered}
\end{equation}
As a result:
\begin{equation}
\p_t\mathcal{A}= {\mathcal{A}}^* \,\p_\alpha\mathcal{A} + \mathcal{A} \,\p_\beta\mathcal{A},
\end{equation}
and therefore ${\cal A}={\cal A}^*=0$ is indeed invariant under the equations of motion. Finally, we notice that the intersection of ${\cal A}=0$ surface and the classical plane $\chi=0$ is given by the condition $\mathcal{A}(\alpha,\bar\alpha)=0$, which coincides with that for the stationary points of the classical equation (\ref{alpha}). This shows that the two-dimensional manifold of dynamical trajectories, departing from the classical stationary points, coincides with the Riemann surface \eqref{Eq:Azero}.

First, notice that fixing $\beta$ on the Riemann surface, allows one to uniquely determine 
\begin{equation}
    \label{eq:alpha-of-beta}
 \alpha(\beta) =\frac{i\tilde\Delta \beta}{G-\tilde \eta \beta^2}.   
\end{equation}
 Substituting this into the equation of motion \eqref{EqAlpha1}: \mbox{ $\dot \beta=(\alpha(\beta) -\bar \beta)\partial_\alpha {\cal A}|_{\alpha=\alpha(\beta)}$,}  one finds
\begin{equation}
                        \label{beta}
 \p_t \beta=i\tilde{\Delta}\beta -   G\bar{\beta}+\tilde{\eta} \bar{\beta}\beta^2.
\end{equation}
Comparing it with the classical (i.e. $\chi=0$) equation of motion (\ref{alpha}), one notices that it coincides with the time-reversed version of the classical equation for $\alpha^*(t)$. This shows that excitation trajectories, following the Riemann surface \eqref{Eq:Azero}, are time-reversed versions of classical relaxation trajectories in the $\chi=0$ plane. This is the origin of HTRS \cite{Carde2025} and closely resembles the situation in {\it equilibrium} stochastic systems. Yet, there the large deviation function (\ref{eq:large-deviation-class}) is fully analytic, while its HTRS quantum analog (\ref{eq:large-deviation-quant}) is not. We now proceed to explain this phenomenon.

The fact that the trajectories follow the Riemann surface allows one to determine the actions, \eqref{effpot}, without knowing detailed shapes of the trajectories. Indeed, the integration contour may be deformed and the action depends only on initial and final points:
\begin{equation}\label{effpot-1}
S(\alpha,\bar{\alpha}) = 2(\bar\alpha\alpha - \bar\alpha_0\alpha_0)-2\!\int\limits_{\alpha_0}^\alpha \! \beta(\alpha)\, d\alpha 
-2\!\int\limits_{\bar \alpha_0}^{\bar\alpha} \! \bar\beta(\bar\alpha)\, d\bar\alpha. 
\end{equation}
The crucial observation is that, while $\alpha(\beta)$ is a single-valued function, this is {\it not} the case regarding its inverse $\beta(\alpha)$.  Indeed, resolving the Riemann surface equation (\ref{Eq:Azero}) for a fixed $\alpha$, one finds two solutions
\begin{equation}\label{betapm}
\begin{gathered}
\beta_{\pm}(\alpha) = 
-i\sqrt{\frac{G}{\tilde{\eta}}}\,\, 
\frac{1  \pm \sqrt{1 -  (\alpha/\alpha_b)^2}}
{(\alpha/\alpha_b)}=i\frac{\delta}{2}\frac{d\phi^{(\pm)}_{0}(\alpha/\alpha_b)}{d\alpha},
\end{gathered}
\end{equation}
where branching points $\pm\alpha_b$ and an effective (complex) inverse Planck constant $\delta$ are defined in Section \ref{SectWigFun}. Thus, the quantum field component $\beta(\alpha)$ (\ref{betapm}) is a multivalued function. This is a consequence  of the trajectory manifold -- the genus zero Riemann surface (\ref{Eq:Azero}) -- being built by gluing two sheets of the physical $\alpha$-phase-space. Such gluing occurs along the cuts (in each sheet) that go between the two branching points, $\pm \alpha_b$ (passing through infinity). Therefore, for each observation point, $\alpha$, there are two possible quantum counterparts, $\beta_\pm(\alpha)$, and correspondingly two distinct trajectories leading to $(\alpha, \beta_-(\alpha))$,  and to $(\alpha, \beta_+(\alpha))$. They accumulate two distinct actions \eqref{effpot-1}:
\begin{equation}\label{EffPotAll}
\begin{gathered}
S_{\pm}(\alpha,\bar{\alpha})=
2\bar{\alpha}\alpha+
2\,\text{Im}\left\{\delta \, \phi_{0}^{(\pm)}(\alpha/\alpha_b)\right\}-C_0,
\end{gathered}
\end{equation}
where 
\begin{equation}
    C_0=2\bar{\alpha}_0\alpha_0+
2\,\text{Im}\left\{  \delta\, \phi_{0}^{(-)}(\alpha_0/\alpha_b)\right\}. 
\end{equation}
Here $\alpha_0=0$ in the stable regime, $G\leq |\tilde\Delta|$, and is given by Eqs.~\eqref{averagen}, \eqref{eq:theta} in the bistable case, $G>|\tilde\Delta|$.

These two solutions correspond to the two sheets that cover the complex $\alpha$-plane. To visualize  this idea, we plot real-time trajectories, spanning the fluctuation Riemann surface, ${\cal A}=0$, obtained by numerical solutions of the equations of motion \eqref{EqAlpha1}, Fig.   \ref{FigPhasePortriet}. The first row, a-c, shows the stable phase, where the only stationary classical solution of (\ref{alpha}) is $\alpha_0=0$, while the second row, d-f, shows the bistable case, with three stationary points, shown as red dots.  Figures \ref{FigPhasePortriet}a,d show the single copy of the complex $\beta^*$-plane. The trajectories, found as solutions of Eq.~(\ref{beta}), emanate from the classical stable points $\beta^*_0=0$ (a), and $\pm \beta^*_0 = \pm\alpha_0$ (d), and run away towards infinity. They are time-reversed versions of the classical relaxation dynamics, described by Eq.~(\ref{alpha}). However, in this case, each physical phase-space point, $\alpha$, has {\it two} images (\ref{betapm}) on the $\beta$-plane, shown in orange and green.   
Figures \ref{FigPhasePortriet}b,c and e,f show the corresponding two copies of the complex $\alpha$-plane, with the cut (shown as dashed blue line) chosen along the real axis. This particular cut translates into the circle with the radius $\sqrt{G/|\tilde \eta|}$ on the $\beta$-plane, such that the interior of the circle corresponds to $\beta_-(\alpha)$ solution, while the exterior corresponds to $\beta_+(\alpha)$ (a different choice of the cut on the $\alpha$-sheets translates into another closed curve encircling the origin on the $\beta$-plane).  All dynamical trajectories originate at the classical stationary point $\alpha =\beta^* =0$ (Fig. \ref{FigPhasePortriet}a-c) and $\alpha=\beta^*=\pm\alpha_0$ (Fig. \ref{FigPhasePortriet}d-f). They all run towards $\beta\to \infty$, smoothly crossing from $\beta_-$ to $\beta_+$. 
On the $\alpha$-plane, crossing the cut means reappearing on the second sheet. As a result, the orange trajectory stays entirely on the first (minus) $\alpha$-sheet, while the green one originates on the first sheet, hits the cut and continues to the second (plus) $\alpha$-sheet. 
 The actions \eqref{effpot-1}, accumulated along these two trajectories, are given by Eq.~\eqref{EffPotAll}. 

In the $\hbar\to 0$ limit, where $|\delta|\gg 1$, one of those two trajectories provides an exponentially dominant contribution to the Wigner function. The switch between the dominant trajectories takes place along the line where the (real) actions \eqref{EffPotAll} are equal. This leads to the anti-Stokes condition \eqref{eq:anti-Stokes}, which specifies the lines on the $\alpha$-plane where the large deviation function \eqref{eq:large-deviation-quant}  exhibits discontinuity in its first derivative.     

The problem with the instanton trajectories analysis presented above, is that $S_+(\alpha,\bar\alpha)$ action is logarithmically divergent at $\alpha\to 0$, cf. Eq.~\eqref{Eq:phiPapp}. This is clearly nonphysical. The origin of this phenomenon is the same as the formal presence of exponentially growing solutions of the Schrödinger equation in the classically forbidden regions. Such solutions should be disregarded as being nonphysical. In the WKB analysis of, e.g. Airy equation, this manifests itself  through the fact that for some region of (complexified) coordinate, the integration contour can pass only through one of the two saddle points. In the other region (which includes classically allowed coordinates), both saddle points contribute to the WKB wave function. These two regions are separated by  Stokes lines, which emanate from the classical turning points into the complex coordinate plane.    

\subsection{Stokes Phenomenon}
\label{sec:Stokes}

\begin{figure}
\includegraphics[width=0.470\textwidth]{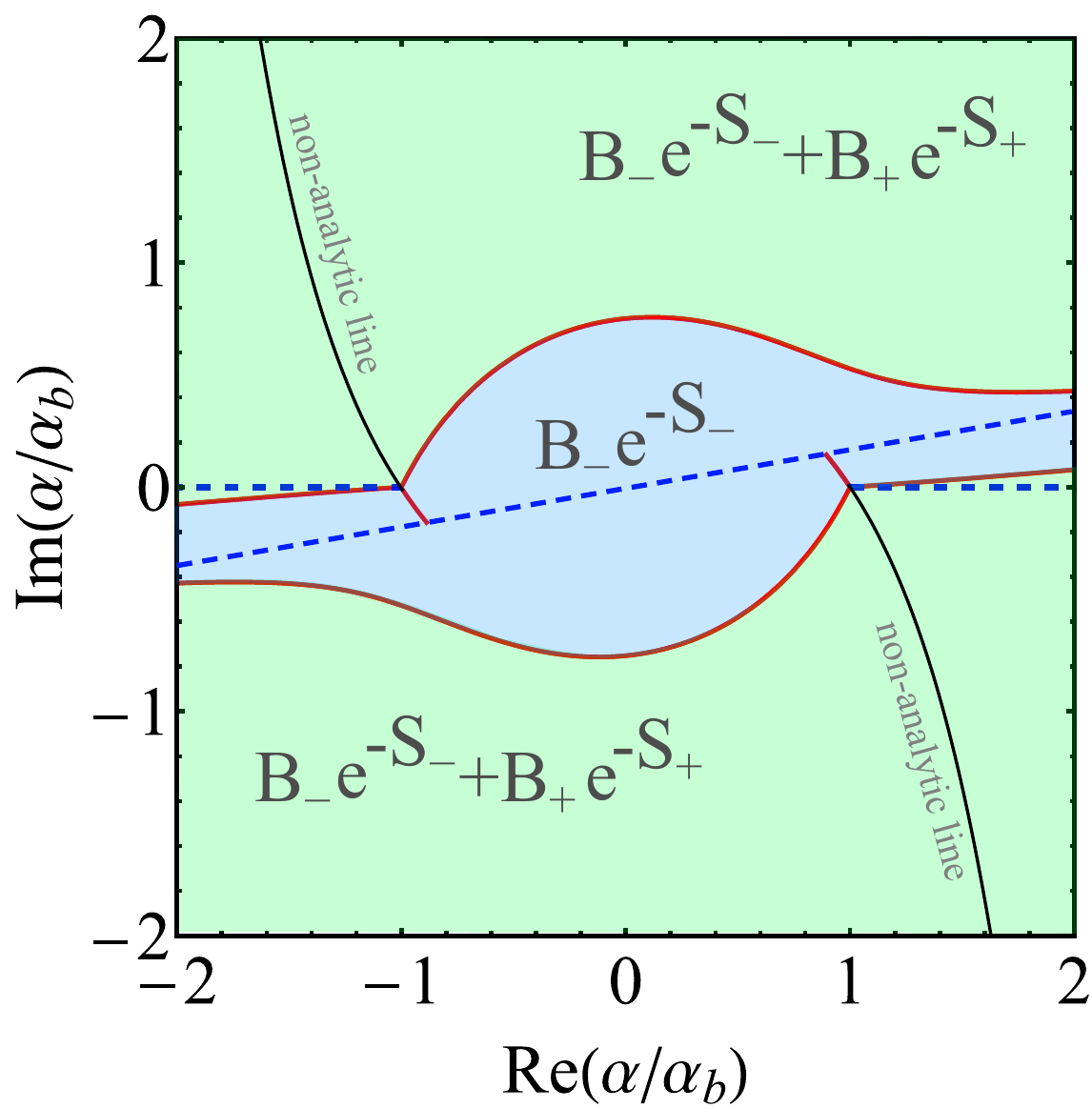}
\caption{
Stokes lines (red) on the complex $\alpha$ plane, as defined by Eq.~(\ref{eq:Stokes}). Within the inner Stokes region (blue), only the instanton with $S_-$ action contributes to the Wigner function. Within the outer Stokes region (green), both instantons with $S_\pm$ contribute. The two actions equalize along the  anti-Stokes lines (black), given by Eq.~(\ref{eq:anti-Stokes}).
The large deviation function is non-analytic along these lines.  One possible choice of branch cuts of the function $\phi^{(\pm)}_0(\alpha/\alpha_b)$ are depicted as blue dashed lines. The branch cut of the logarithm $\ln(\alpha^2)$ in $S_+$ should be chosen within the inner Stokes region. The system parameters are set to $G=2, \Delta=2, \kappa=0.4, \eta=1, U=0.1, \kappa_\phi=0$.}
\label{FigStokesAllTwo}
\end{figure}

Here we show that the same Stokes phenomenon takes place for the instanton treatment of large deviations. Namely,  the complex $\alpha$-phase-space is split by the Stokes lines into a region where only $S_-(\alpha,\bar\alpha)$ contributes to the Wigner function \eqref{WigAct}, and a region where both $S_\pm(\alpha,\bar\alpha)$ should be kept, as shown in Fig. \ref{FigStokesAllTwo}. The former region includes $\alpha =0$, thus avoiding the singularity of $S_+(0,0)$, while the latter includes switching ``mountain range'' anti-Stokes lines. At Stokes lines, the pre-exponential factor $B_+$ abruptly changes to zero. The Wigner function $W_0(\alpha,\bar\alpha)$ is still an entire function without any discontinuities at the Stokes lines. This is consistent because at Stokes lines $e^{-S_+}\ll e^{-S_-}$, and an abrupt change in $B_+$ is compensated by an exponentially small (non-perturbative)  change in $B_-$ \cite{witten2010}.

The functional integral representation of the Wigner function (\ref{eq:funct-integral}) is a shorthand notation for multiple integrals labeled by discretized time steps. Imagine all those integrations being performed, apart from the very last, $N$-th, time step. In this last step, the classical field is fixed at the observation point $(\alpha,\bar\alpha)$ and one is left  with integrations over the quantum field (through the change of variables $\alpha_q\to -\bar\chi\to \beta$): 
\begin{equation}
                \label{eq:last-step}
    W_0(\alpha,\bar\alpha)=e^{-2\bar\alpha\alpha}\iint\! d\beta d\bar\beta\, \, e^{2\beta\alpha+2\bar\beta\bar\alpha -F(\beta,\bar\beta)},
\end{equation}
where, e.g., $2\beta\alpha$ is a part of the discretization of $2\beta\dot\alpha dt  \to  2\beta_N(\alpha_N-\alpha_{N-1})$.  Here, $\alpha_N=\alpha; \beta_N=\beta$ and the integral over $\alpha_{N-1}$ is performed at the previous time step, resulting in a $\beta_N=\beta$-dependent function denoted as $e^{-F(\beta,\bar\beta)}$. In the semiclassical approximation, \mbox{$F(\beta,\bar\beta)=2\int \!dt\, (\alpha\dot\beta +\bar\alpha\dot{\bar\beta})  $,} where the integral is calculated along an instanton trajectory with  fixed $(\beta,\bar\beta)$ as its final point. Here we have employed the fact that ${\cal L}=0$ along such a trajectory.   

\begin{figure}
\includegraphics[width=0.45\textwidth]{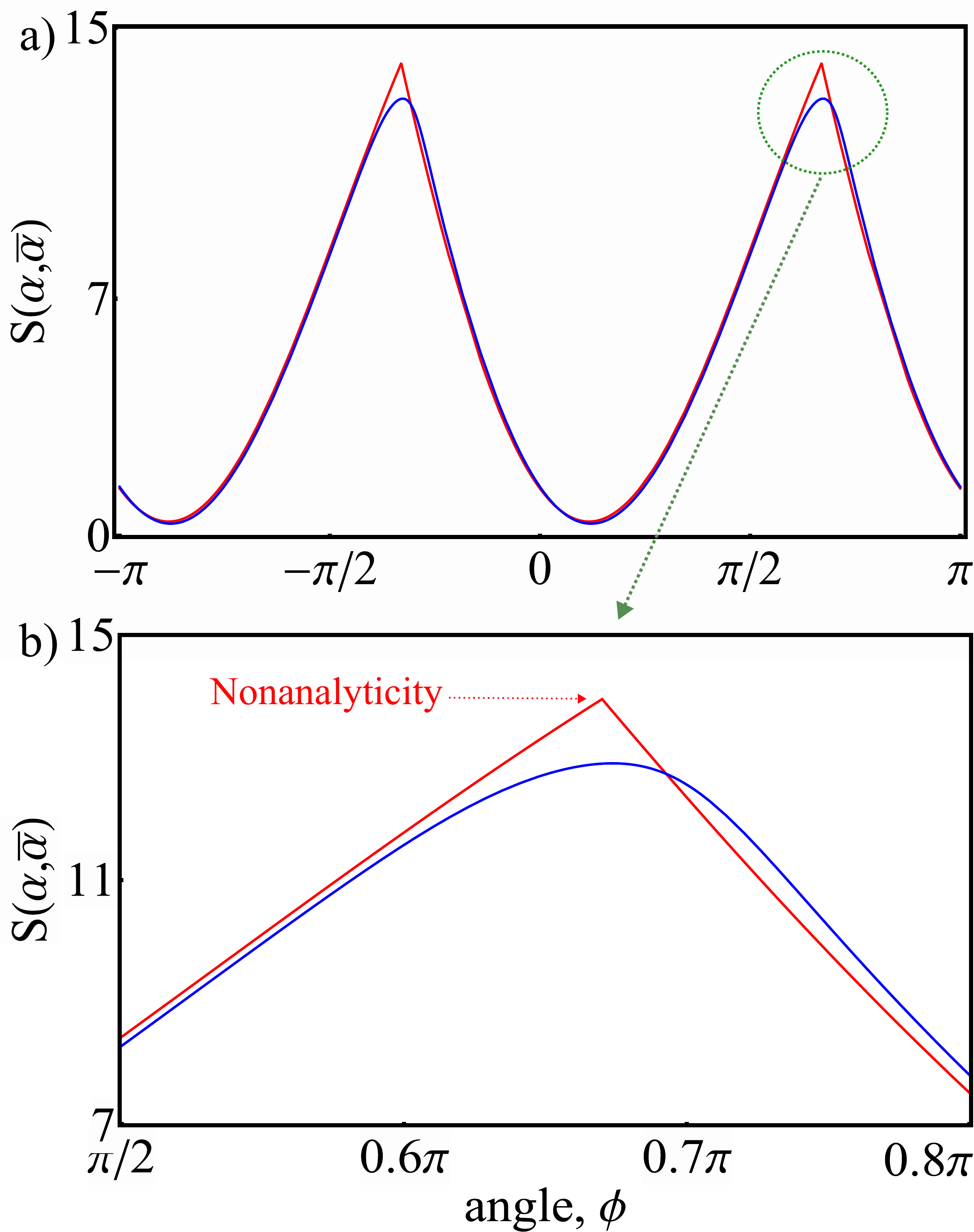}
\caption{(a,b) The large-deviation function from the instanton approach \eqref{LDF1} (red) and the exact solution \eqref{Decomp} -- \eqref{EfPotbase} (blue) vs. the angle $\varphi=\arg(\alpha/\alpha_b)$ for a fixed radius  $|\alpha|=3.5|\alpha_b|$.  The system parameters are set to \mbox{$G=5,\Delta=2$},\mbox{ $\kappa=0.4, \eta=1$}, $U=0.1, \kappa_\phi=0$.}
\label{PotentialSlice}
\end{figure}

In the HTRS case, such a trajectory is {\it unique}, cf. Figs.~\ref{FigPhasePortriet} (a,d) with a holomorphic dependence $\alpha(\beta)$ (cf. Eq. \eqref{eq:alpha-of-beta}). One thus finds $F(\beta,\bar\beta) = F(\beta)+F^*(\bar\beta)$, where $F(\beta)=2\int^\beta \! \alpha(\beta)\,d\beta$. As a result, Eq. \eqref{eq:last-step} dictates the holomorphic form (\ref{Decomp}) of the Wigner function with 
\begin{equation}
                \label{eq:Psi-Fourier}
    \Psi(\alpha)\propto  \int_{\cal C}\! d\beta\,\, e^{2 \beta \alpha-F(\beta)}; \qquad
    F(\beta) = 2\!\int\limits^\beta\frac{i\tilde{\Delta}\beta \, d\beta}{G-\tilde{\eta}\beta^2}, 
\end{equation}
where the integration contour, ${\cal C}$, connects the regions where ${\mathrm{Re} }(2\beta \alpha-F(\beta)) <0$ and passes between the branching points $\beta = \pm \sqrt{G/\tilde \eta}$. One may recognize that this expression is nothing but the Fourier transform approach to  solving Eq.~(\ref{difeq1}). The latter is a quantized version of the Riemann surface condition (\ref{Eq:Azero}): ${\cal A}(\alpha,\partial_\alpha/2)\Psi(\alpha)=0$. Its Fourier transform version is ${\cal A}(-\partial_\beta/2,\beta)e^{-F(\beta)}=0$, where within the accuracy of the semiclassical approximation, we disregard the non-commutativity of the operators. Equation~(\ref{eq:Psi-Fourier}) for $F(\beta)$ is the straightforward solution of this latter {\it first-order} differential equation.  

The saddle point condition for the integral in Eq.~(\ref{eq:Psi-Fourier}) yields $\alpha=i\tilde{\Delta}\beta/(G-\tilde{\eta}\beta^2)$, which coincides, of course, with the Riemann surface condition (\ref{Eq:Azero}). It leads to two saddle points $\beta_{\pm}$, given by Eq.~(\ref{betapm}). The integration contour ${\cal C}$ may pass through one ($\beta_-$) or both of these saddle points, depending on the value of $\alpha$.  These two regimes  are separated on the $\alpha$-plane by the Stokes lines, which emanate from the ``turning points'', $\pm \alpha_b$, defined  by $\beta_-(\pm\alpha_b) =\beta_+(\pm\alpha_b)$. The Stokes lines are determined by the condition \cite{bender1999,witten2010}:
\begin{equation}
    \label{eq:Stokes}
    \mathrm{Re}\left\{ \delta\, \phi_{0}^{(+)}(\alpha/\alpha_b)\right\} = \mathrm{Re}\left\{\delta \, \phi_{0}^{(-)}(\alpha/\alpha_b)\right\},
\end{equation}
where we used the fact that
\begin{equation}\label{Prop1}
2 \beta_\pm(\alpha)\alpha  -F(\beta_\pm(\alpha)) = i\delta\, \phi_{0}^{(\pm)}(\alpha/\alpha_b).
\end{equation}
Figure~\ref{FigStokesAllTwo} shows  the Stokes lines emanating from the two branching (turning) points $\alpha=\pm \alpha_b$. In the inner Stokes region, which contains $\alpha=0$,  only the instanton with a regular action, $S_-(\alpha,\bar\alpha)$, is allowed by the contour integral in \eqref{eq:Psi-Fourier}, while the one with the singular action, $S_+(\alpha,\bar\alpha)$, is absent. In the outer regions, both instantons are present. In the outer region, the switching between the two instantons  occurs along the non-analytic anti-Stokes line determined by Eq. \eqref{eq:anti-Stokes}. 
As a result, the Stokes phenomenon leads to the following expression for the large-deviation function: 
\begin{equation}\label{LDF1}
    S(\alpha,\bar\alpha)=\begin{cases} S_-(\alpha,\bar\alpha); &\alpha \in R_{I},  \\ \min\big\{S_-(\alpha,\bar\alpha),S_+(\alpha,\bar\alpha)\big\}; &\alpha \in R_{O},  \end{cases}
\end{equation}
where $R_I$ and $R_O$ denote the inner and outer Stokes regions, defined through Eq.~(\ref{eq:Stokes}) and shown in Fig.~\ref{FigStokesAllTwo}. 

Figure~\ref{PotentialSlice} shows a comparison between the large deviation function \eqref{LDF1} and the exact solution \eqref{Decomp}--\eqref{EfPotbase}. They are in almost perfect agreement. A slight discrepancy (amplified in panel b) can be observed near the non-analytic point of the large-deviation function \eqref{LDF1}.  The non-analyticity is smeared in the exact solution due to finite quantum fluctuations, as $|\delta|=2.03$ is not too large.

\subsection{Phase-slip rate}
\label{sec:phase-slips}

\begin{figure}
\includegraphics[width=0.420\textwidth]{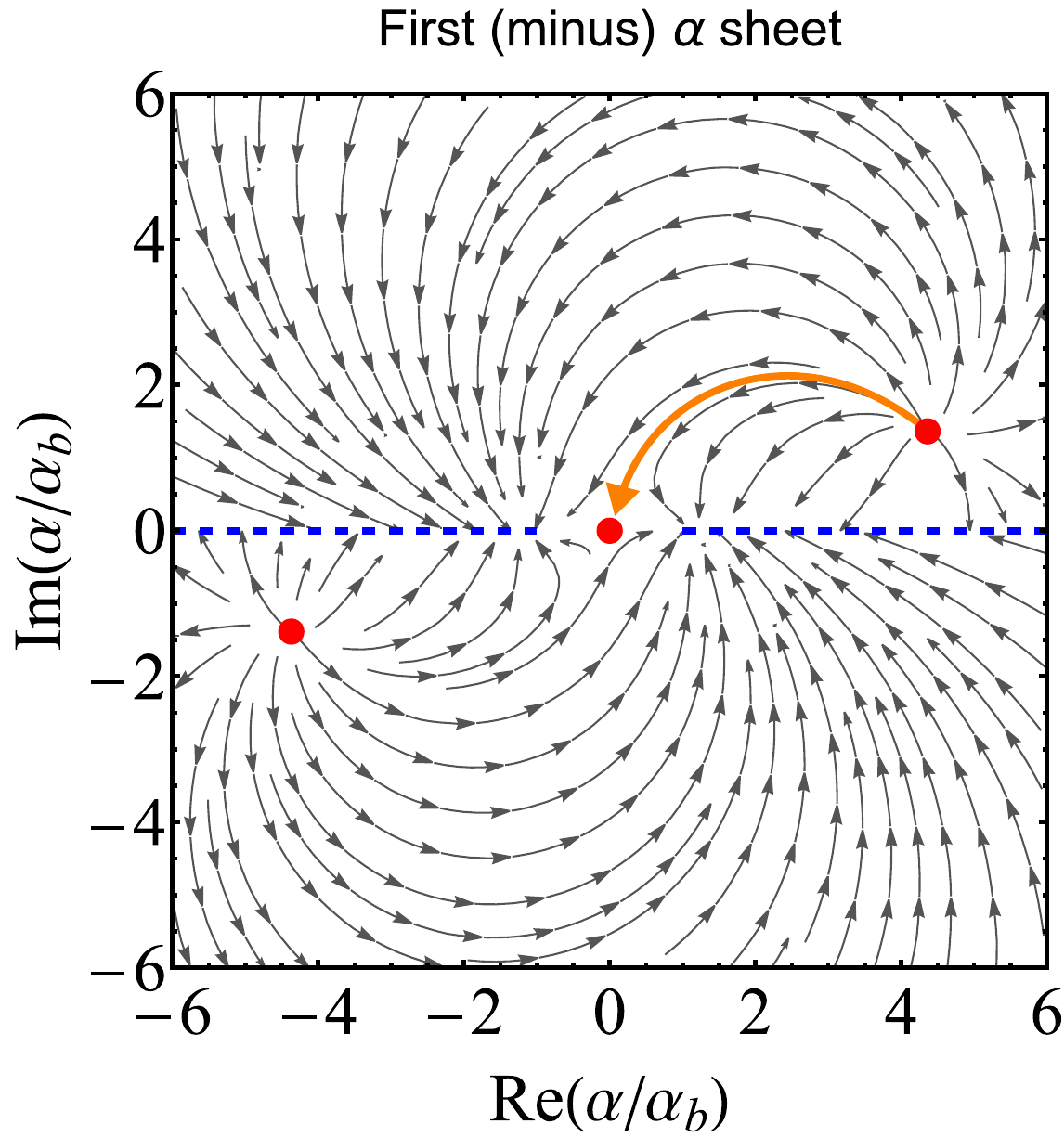}
\caption{The $\alpha$-plane flow and the phase-slip instanton path (orange line) in the bistable regime. The phase-slip path starts at the classical fixed point $\alpha=\alpha_0$ and ends at the saddle point $\alpha=0$ (red dots). From the saddle point to the opposite stationary point, the system evolves classically (i.e., in $\chi=0$ subspace), according to Eq.~(\ref{alpha}), with no action accumulated along this part of the phase-slip path.    The system parameters are set to $G=5, \Delta=2, \kappa=0.4, \eta=1,U=0.1, \kappa_\phi=0$.}
\label{FigInstanton2}
\end{figure}

Here we briefly comment on the quantum phase-slip rate as it follows from the instanton formalism. In the bistable regime, $G>|\tilde{\Delta}|$, the classical dynamics of the system exhibits two stable fixed points, $\alpha=\pm \alpha_0$, separated by an unstable saddle point at $\alpha=0$. Within purely classical dynamics, a system prepared near one of these points would remain in its vicinity indefinitely. However, if the system is initialized near one of the fixed points, quantum fluctuations can eventually flip it to the other. Macroscopically, this process manifests itself as a  $\pi$ phase-slip of the parametrically induced oscillations \cite{Marthaler2007,Dykman1994,Dykman1998,Dykman2007,Dykman2012,Su2025,Frattini2024,Boness2025}. Such quantum phase-slips constitute an ultimate limit for the coherence of quantum devices based on parametric oscillators \cite{Putterman2025,Dubovitskii2024,Dubovitskii2025,Chamberland2022}. Significant progress has been made in recent years in calculating  the phase-slip rate using different approaches, such as the complex P-representation \cite{Kinsler1991,Sun2019}  and estimating the instanton action both numerically \cite{Lee2025,Xiang2026} and analytically \cite{Thompson2022,Sepulcre2026,Carde2025,Thompson2026}. 
Here, we show how to calculate the phase-slip rate using the large deviation function and recover various results from several recent studies  \cite{Carde2025,Sun2019,Thompson2022} within a unified framework.  

As shown in Sec.~\ref{sec:Stokes}, in the vicinity of the saddle point $\alpha=0$, the contribution associated with the $S_{+}$ action is excluded by the Stokes phenomenon.  Therefore, the phase-slip rate $\Gamma$ is governed by the trajectory belonging to the $S_{-}$ branch:
\begin{equation}
\Gamma \propto
\exp\left[-S_{-}(0,0)+S_{-}(\alpha_0,\bar{\alpha}_0)\right],
\end{equation}
where $S_{-}(\alpha,\bar{\alpha})$ is given by Eq.~\eqref{EffPotAll}. This leads to the following expression:  
\begin{equation}\label{lnGamma1}
\begin{split}
-\ln\Gamma=-2\bar{\alpha}_0\alpha_0+2 \mathrm{Im}\left[\delta\left(\phi_0^{(-)
   }(0)-\phi_0^{(-)}\left(\frac{\alpha_{0}}{\alpha_b}\right)\right)\right].
\end{split}
\end{equation}
Here,  $\phi_0^{(-)}(z)$ is given by Eq.~\eqref{Eq:phiMapp} and $\delta$ is the complex normalized detuning defined in Section \ref{SectWigFun}. After simple algebra, one obtains the phase-slip rate straightforwardly:
\begin{equation}\label{LnGamma2}
    -\ln{\Gamma}=2n_0+2\mathrm{Re}\left( {\tilde{\Delta}\over \tilde{\eta}}\right)\left( 2\theta_0 -{\pi \over 2}\right)+ 2\mathrm{Im}\left(\frac{\tilde{\Delta}}{\tilde{\eta}} \ln\left(\frac{G}{\tilde{\Delta}}\right) \right),
\end{equation}
where $n_0$ and $\theta_0$ are given by Eqs. \eqref{averagen}-\eqref{eq:theta} and the complex logarithm in Eq. \eqref{LnGamma2} is taken with the principal branch cut. 

Recently, the phase-slip rate for this system was calculated using the framework of the complex P-representation within the Fokker-Planck approach \cite{Sun2019} and the instanton formalism \cite{Carde2025}. Here we briefly demonstrate the equivalence of the two approaches.  The complex P-distribution is defined through the following representation of the reduced density matrix:  \cite{Bartolo2016,Drummond1980}
\begin{equation}
\hat{\rho}(t)=\int_\mathcal{C}\!\!d\beta \int_{\mathcal{C}'}\!\!d\bar{\beta}\, \frac{\ket{\bar{\beta}}\bra{\beta^*}}{\braket{\beta^*|\bar{\beta}}}\, P(\beta,\bar{\beta};t).
\end{equation}
It is related to the Wigner function \eqref{WignerExact} through the integral transformation 
\begin{equation}\label{WignP}
W(\alpha,\bar\alpha;t)=\mathcal{N}\iint d\beta d\bar\beta\,\, e^{-2(\alpha-\bar{\beta})(\bar\alpha-\beta)}\,P(\beta,\bar\beta;t).
\end{equation}
Comparison with Eq.~\eqref{eq:last-step} yields the stationary P-distribution function in the semiclassical limit ($|\delta|\gg1$):
\begin{equation}\label{Pfuntion}
P_0(\beta,\bar{\beta}) \propto \exp\left\{2\bar{\beta} \beta- F(\beta)-F^*(\bar\beta)\right\},
\end{equation}
where $F(\beta)$ is given by Eq.~\eqref{eq:Psi-Fourier}. The same stationary P-distribution function was  found by solving the stationary Fokker-Planck equation \cite{Bartolo2016}. As has been shown in several papers \cite{Kinsler1991,Sun2019,Carde2025}, the stationary P-distribution provides the phase-slip rate as: 
\begin{equation}\label{GammaActionFinal}
-\ln\Gamma = -\ln \left(\frac{P_0(\beta=\bar{\alpha}_0,\bar{\beta}=\alpha_0)}{P_0(\beta=0,\bar{\beta}=0)}\right).
\end{equation}
Employing Eqs.~\eqref{Pfuntion} and \eqref{Prop1}, one finds that the phase-slip rate \eqref{GammaActionFinal} coincides with Eq. \eqref{lnGamma1}. 

\begin{figure}
\includegraphics[width=0.45\textwidth]{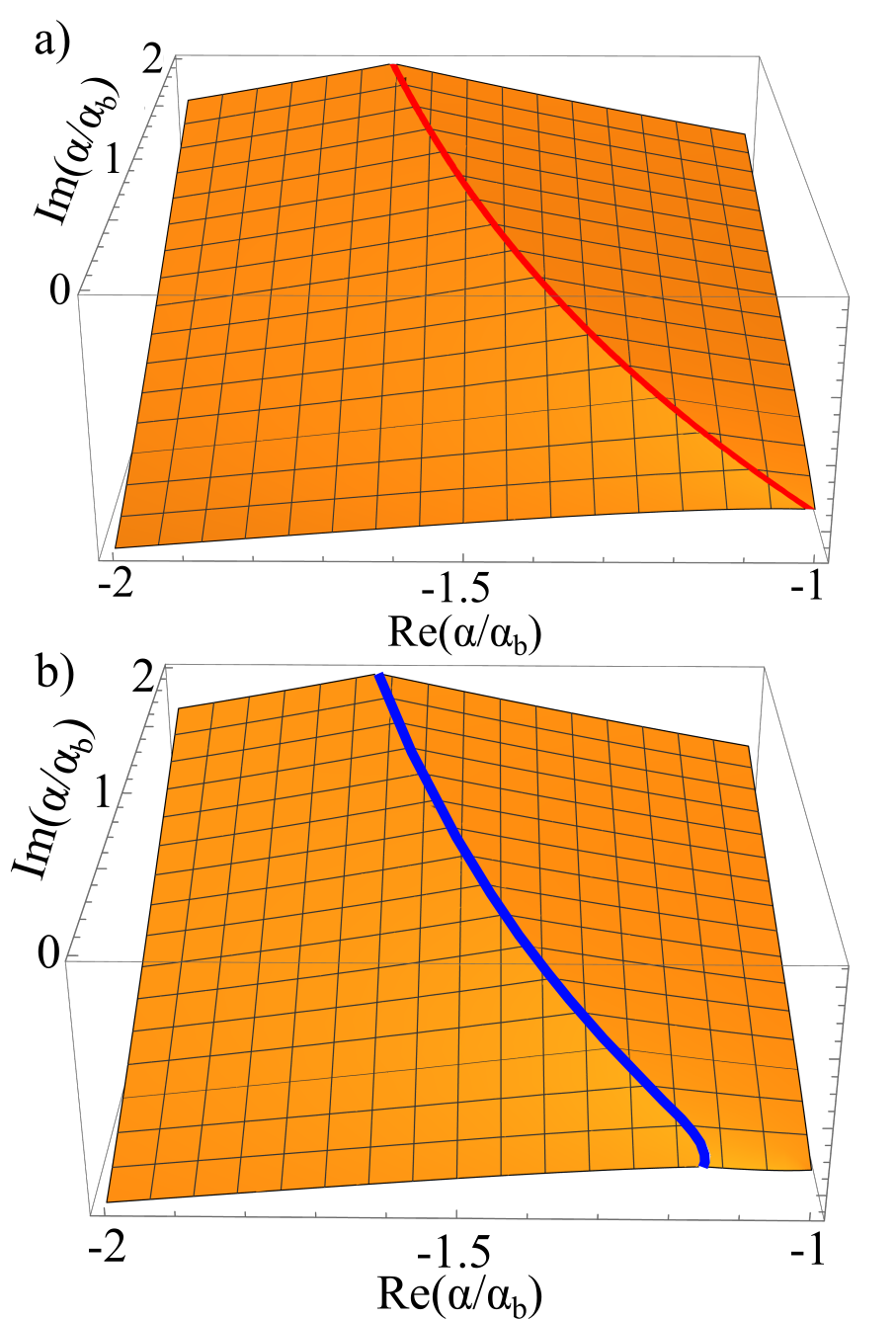}
\caption{(a-b) The large deviation function $S(\alpha,\bar\alpha)=\min\{S_-(\alpha,\bar{\alpha}),S_+(\alpha,\bar{\alpha})\}$ vs. phase-space coordinates $\rm{Re}(\alpha)$ and $\rm{Im}(\alpha)$. It is calculated numerically for zero dephasing rate $\kappa_{\phi}=0$ (a) and finite dephasing $\kappa_{\phi}=0.03$ (b) using Eq. \eqref{overallAct}. Red and blue curves correspond to nonanalytic lines, where two instanton actions are equal $  S_{-}(\alpha,\bar{\alpha})=S_{+}(\alpha,\bar{\alpha})$. The system parameters are set to \mbox{$G=5,\Delta=2$},\mbox{ $\kappa=0.4, \eta=1$}, $U=0.1$.}
\label{nonanalyticline}
\end{figure}

\subsection{Hidden time-reversal symmetry breaking}

The previous sections considered instanton dynamics in HTRS systems, where instantons are confined to the Riemann surface, allowing one to obtain analytical solutions. In this section,  dephasing processes characterized by a rate $\kappa_{\phi}$ are taken into account. The presence of the dephasing violates HTRS and causes instanton trajectories to deviate from the Riemann surface. This complicates the analytic treatment. Some progress can be made in first order perturbation theory, treating $\kappa_{\phi}$ as a small parameter. 

The dephasing term changes the equations of motion, making the trajectories deviate from their HTRS  paths along the Riemann surface. It also adds an extra term, $\propto \kappa_\phi$, to the action, Eq.~(\ref{Li1}). Since the bare trajectories extremize the HTRS action, there is no change to the HTRS action in the first order in $\kappa_\phi$. As a result, the first order correction comes entirely from the $\propto \kappa_\phi$ term, 
\begin{equation}\label{overallAct}
    S_{\pm}(\alpha,\bar{\alpha})= S_{\pm}^{(0)}(\alpha,\bar{\alpha})+\kappa_\phi \,  S^{(\phi)}_{\pm}(\alpha,\bar{\alpha}),
\end{equation}
\begin{equation}\label{EqDef}
    S^{(\phi)}_{\pm}(\alpha,\bar{\alpha})=-8\!\int\! dt \,\,  \text{Im}^2\{\alpha_{\pm}(t)\beta_{\pm}(t)\}.
\end{equation}
which is calculated along the {\it bare} HTRS trajectories,  ($\alpha_{\pm}(t),\beta_{\pm}(t)$), and the unperturbed action $S_{\pm}^{(0)}$ is given by \eqref{EffPotAll}. From Eq. \eqref{EqDef} it follows that  dephasing reduces the overall action \eqref{overallAct} (at least in the first order). It thus increases the phase-slip rate by virtue of adding an extra source of noise. 

The extra action (\ref{EqDef}) shifts the location of the non-analytic line within the physical $\alpha$-phase-space. This is illustrated in Fig. \ref{nonanalyticline}, where we calculated (\ref{EqDef}) through numerical integration of the bare trajectories, and plotted the large deviation function (\ref{LDF1}). One may notice  that the largest shift of the non-analytic line occurs near the bare branching point. This is because $S_-(\alpha,\bar\alpha)-S_+(\alpha,\bar\alpha)$ is very flat there, and therefore its zero is easily shifted by a small perturbation.  
 
 The message from this consideration is that the non-analytic lines are perturbatively stable against small HTRS-breaking  perturbations. They are shifted around but do not disappear. One cannot exclude, however, non-perturbative effects, such as, e.g., appearance of extra non-analytic lines in  distant parts of the phase space.

\section{Conclusions}
\label{sec:conclusions}

In this work, we investigated stationary states of driven open quantum systems. Specifically, we focused on probabilities of large atypical quantum fluctuations - the so-called rare events. 
Our main result is that the large deviation function is  non-analytic in phase-space coordinates characterizing the rare event. Namely, there are lines across the phase space, along which all derivatives, starting from the first, of the large deviation function are discontinuous. Such lines emanate from  special ``turning'' points, which are located at a finite distance from classical stationary states. The non-analyticity in the large deviations is present even though the probability of typical, small-enough, fluctuations is perfectly smooth and analytic.   Several remarks are in order:

(i) Details of the non-analytic structure are only weakly sensitive to a strength of system--bath coupling. It  persists even in the limit where such coupling is taken to zero. Yet, it is crucial that there is a coupling to a bath -- no matter how weak. Without such a coupling, the only steady state, the system can reach, is  trivial infinite temperature state. Once a weak coupling is present,  the system reaches a non-trivial steady state, which is {\it very} different from a ground state of an effective rotation-wave Hamiltonian -- no matter how small the bath-induced dissipation rates are (the latter do affect how fast the steady state is reached). 
For small fluctuations this observation manifests itself in {\it quantum heating} \cite{Dykman1988,Dykman2011} -- an effective finite temperature 
in Floquet basis, even when the bath temperature is zero. For larger fluctuations it results in {\it quantum activation} \cite{QuantumActivation} of phase-slips, and for even larger ones -- in the non-analytic structure, discussed here. 

(ii) A somewhat similar non-analyticity is known 
to show up in the large deviation functions of non-equilibrium {\it classical} stochastic systems \cite{Freidlin1998,Dembo2009,Graham1984a,Graham1984b,Graham1985,Graham1983,Dykman1994}.  However, the details are rather different. In the classical context, the necessary condition for 
the non-analyticty to emerge is breaking of time-reversal symmetry between relaxation and activation trajectories, associated with the detailed balance condition. This may lead to a folding catastrophe of the activation trajectories Lagrangian manifold.  It results in caustics in the projection of this manifold onto the physical phase space. Between such caustics there are, in general, three trajectories leading to a given phase-space location.  
In quantum case, the non-analytic structure appears even in the presence of (hidden) time-reversal symmetry   \cite{Roberts2021}. In this case the trajectories manifold follows the Riemann surface. The latter does not fold on itself, but is rather glued out of two sheets of the (complex) phase space. Consequently there are two (not three) trajectories leading to {\it every} phase-space location (not only regions between caustics).  

(iii) Another related feature, which, to the best of our knowledge, has not been seen in the classical context, is the Stokes phenomenon. 
Since {\it any} point in the phase space can be reached by the two trajectories,  one should be aware of the fact that both corresponding actions cannot be regular in the entire phase-space. 
One should be careful thus to exclude trajectories with  a singular action in some parts of the phase-space. This is taken care of by the Stokes phenomenon. It originates from the fact that, for some parameters, the (functional) integration contour(s) can't pass through both stationary trajectories, and only one of them (with a non-singular action) contributes to the probability of the rare event.
It requires a future work to understand how this phenomenon plays out outside of the HTRS scenario. 

Other questions left to future studies include how the Riemann surface topology of the activation manifold affects bifurcation phase transition in extended systems (e.g., a chain of parametrically driven oscillators \cite{vicentini2018}). Does the holomorphic structure (\ref{Li2}) of the action  survive upon renormalization? If indeed, does this symmetry affects universality class \cite{sieberer2025} of the dynamical $Z_2$ symmetry breaking bifurcation transition?

\begin{acknowledgments}
We are grateful to Mark Dykman, Nikita Nekrasov,  and Foster Thompson for valuable discussions. We also thank Igor Burmistrov, Danil Bugakov and Alexander McDonald for valuable feedback on our work. S.O. Potashin thanks the Russian Science Foundation (Project No. 25-12-00212) for financial support of the theoretical study of the Wigner function. V. Yu. Mylnikov acknowledges the support of numerical simulations and the instanton approach by the Foundation for the Advancement of Theoretical Physics and Mathematics “BASIS.” 
\end{acknowledgments}

\appendix

\section{WKB calculation of the Wigner function}\label{AppWKB}

This section provides a detailed calculation of the function $\Psi(\alpha)$ using the WKB method. One first rewrites  Eq. \eqref{difeq1} in terms of a normalized  variable $z=\alpha/\alpha_b$, where $\alpha_b=\tilde{\Delta}/(2\sqrt{G\tilde{\eta}})$:
\begin{equation}\label{difeq11}
  z \Psi''(z)+2i\delta\Psi'(z)-\delta^2 z \Psi(z)=0,
\end{equation}
where the normalized complex detuning $\delta=\tilde{\Delta}/\tilde{\eta}$,  is a large parameter, $|\delta|\gg1$. To apply WKB approximation to Eq. \eqref{difeq11}, one seeks a solution in the exponential form $\Psi(z)= \exp\left\{i\phi(z)\right\}$, which yields for $\phi(z)$: 
\begin{equation}\label{phi01}
   - z(\phi')^2+ iz \phi''-2\delta\phi'-\delta^2 z = 0.
\end{equation}
Employing the large parameter, one may look for a solution in the form  $\phi(z)\approx\delta\phi_0(z)+\phi_1(z) +O(\delta^{-1})$. Substituting this expansion into Eq. \eqref{phi01} gives:
\begin{equation}\label{phi02}
    \delta^2\bigl[-z(\phi_0')^2-2\phi_0'- z\bigr]
    +\delta \bigl[-2 \phi_1' -2 z \phi_0' \phi_1' +i z \phi_0''\bigr] = 0.
\end{equation}
Setting the term $\sim \delta^2$ to zero yields a quadratic equation for $\phi_0'$, which is solved as: 
\begin{equation}\label{derphi}
    \phi_{0}^{(\pm)}{'}(z)=-\frac{1\pm\sqrt{1- z^2}}{z}.
\end{equation}
Integrating these expressions results in: 
\begin{equation}\label{phiMapp}
    \phi_{0}^{(-)}(z)= 
 \sqrt{1- z ^2}
- \ln  \left(1+\sqrt{1- z^2}\right),
\end{equation}
\begin{equation}\label{phiPapp}
 \phi_{0}^{(+)}(z)= -\phi_{0}^{(-)}(z)- 
\ln(z^2).
\end{equation}
Notice that $\phi_{0}^{(-)}(z)$ is regular near $z=0$, while $\phi_{0}^{(+)}(z)$ is singular due to the logarithmic term which comes from the simple pole of \eqref{derphi}. 

Next, let us calculate the function $\phi_{1}(z)$. The equation for $\phi_{1}(z)$ follows from the $\sim \delta$ term in Eq. \eqref{phi02}: 
\begin{equation}\label{phi1}
   \phi_{1}{'}(z) = \frac{iz\,\phi_{0}{''}(z)}{2\bigl(1 + z   \phi_{0}{'}(z)\bigr)}.
\end{equation}
Substituting \eqref{phiMapp} and \eqref{phiPapp} into Eq.~\eqref{phi1} and integrating over $z$, one obtains 
\begin{equation}
    \phi_{1}^{(-)}(z)=\frac{i}{2}\ln\left[\frac{\sqrt{1-z^2}}{1+\sqrt{1-z^2}}\right]-\frac{\pi}{4}.
\end{equation}
\begin{equation}
\begin{gathered}
    \phi_{1}^{(+)}(z)=\frac{i}{2}\ln\left[\sqrt{1-z^2}+ 1-z^2 \right]-\frac{i}{2}\ln\left(z^2\right)+\frac{\pi}{4}.
\end{gathered}
\end{equation}
Introducing  complex amplitudes $A_{\pm}(z)$ related to the functions $\phi_{1}^{(\pm)}(z)$ as:
\begin{equation}\label{Apm}
A_{\pm}(z)=\exp{\left\{i\phi_{1}^{(\pm)}(z)\right\}},
\end{equation}
one obtains  the two WKB solutions for the function $\Psi(z)$, given by
\begin{equation}\label{PsiWkb}
\Psi_\pm(z) = A_{\pm}(z)\exp \left\{i\delta\, \phi_{0}^{(\pm)}(z)\right\}.
\end{equation}

Near the branching point $z=1$ of the square root function, the WKB solutions are as follows: 
\begin{equation}\label{SolZ}
\begin{gathered}
\Psi_\pm(z) \approx
\frac{1}{\left(2(1-z)\right)^{1/4}} \\
\times
\exp\left\{
\pm i\delta\frac{2\sqrt{2}}{3}(1-z)^{3/2}
+i\delta(1-z)\pm i\frac{\pi}{4}
\right\},
\end{gathered}
\end{equation}
with a similar expressions near $z=-1$. 
There are two types of branch cuts of the WKB solutions \eqref{PsiWkb}, shown in Fig. \ref{FigStokesAllTwo}. The first cut arises from the square root $\sqrt{1-z^2}$. We take its principal branch along $(-\infty, -1] \cup [1, \infty)$. The second cut arises from the logarithmic term $\ln z^2$. Although the principal branch cut for $\ln z^2$ is conventionally chosen as the imaginary axis $(-i\infty, i \infty)$, here, it is convenient to deform the logarithmic branch cut along an arbitrary line that lies within the inner Stokes region (see Appendix \ref{AppComPlWKB}). The exact form of the cut is irrelevant, as it affects only the singular WKB solution $\Psi_+$, which is absent inside the inner Stokes region.

\section{Stokes phenomenon near the branching point}
\label{AppComPlWKB}
\begin{figure}
\includegraphics[width=0.420\textwidth]{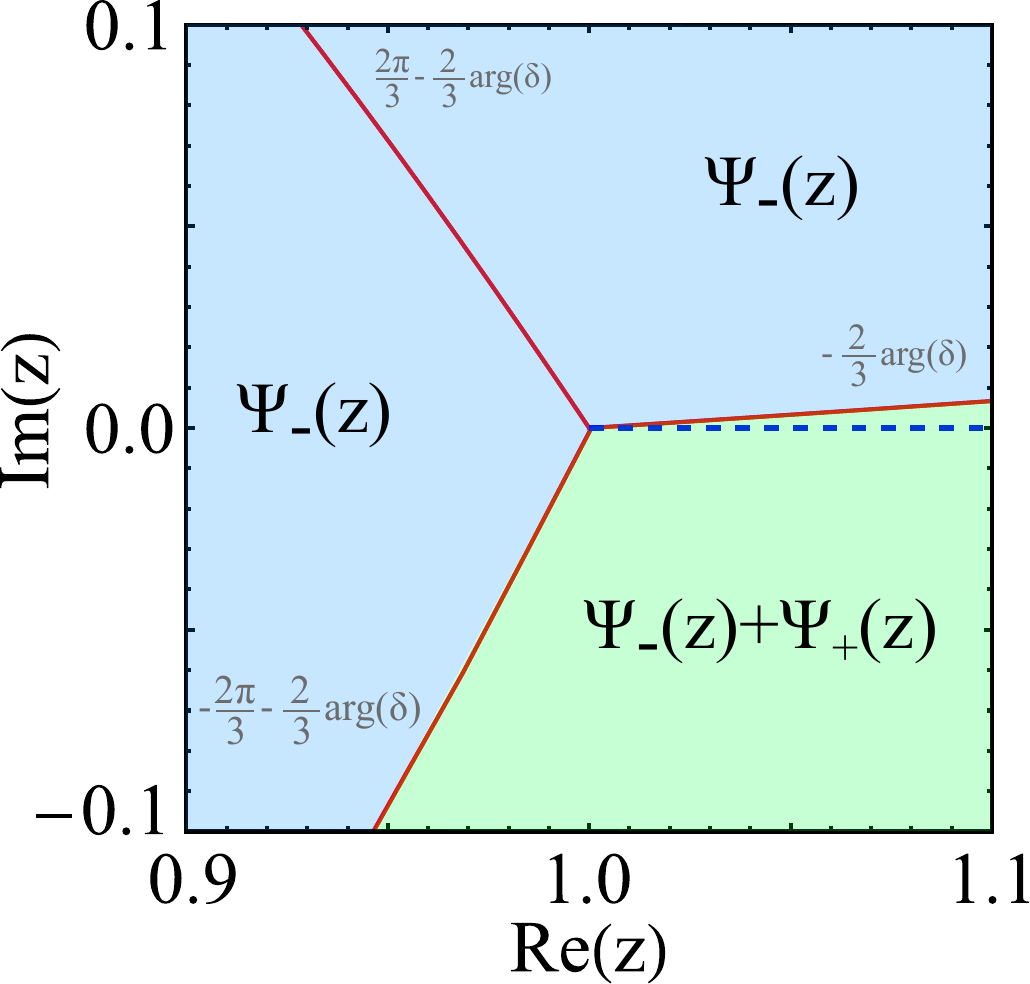}
\caption{
Stokes regions near the  turning point $z=1$. Within the inner Stokes region (blue), there is only one WKB solution $\Psi_-$. Within the outer Stokes region (green), the solution  is the sum of two WKB terms $\Psi_-+\Psi_+$. The Stokes lines are shown in red. A possible branch cut is shown in blue dashed line.}
\label{StokesBrP}
\end{figure}

Here the origin of the Stokes phenomenon for the WKB wave function is discussed using its integral representation \eqref{eq:Psi-Fourier}:
\begin{equation}\label{psiint1}
    \Psi(\alpha)\propto  \int_{\cal C}\! d\beta\,\, \exp\left\{2 \beta \alpha+i\, \frac{\tilde{\Delta}}{\tilde{\eta}}\, \ln \big(G-\tilde{\eta}\beta^2\big)\right\}. 
\end{equation}
It is convenient to consider $\alpha$ near the branching points $\pm\alpha_b$. One can focus on the right branching point because the parity symmetry $\alpha\rightarrow-\alpha$ makes them equivalent. From Eq. \eqref{betapm} it follows that the quantum variable can be rescaled as follows:
\begin{equation}\label{changeofv}
    \beta=-i\sqrt{G/\tilde{\eta}}\, (1+y), 
\end{equation}
where  $|y|\ll 1$ near the branching point. Substituting \eqref{changeofv} into \eqref{psiint1} and expanding the logarithm to the third order in $y$, yields:
\begin{equation}\label{psiint2}
    \Psi(z)\propto  e^{-i\delta z}\int_{\cal C}\! dy\,\, \exp\left\{i \delta y\,(1-z)-i\delta\, \frac{y^3}{6}\right\}, 
\end{equation}
where the rescaled variable $z=\alpha/\alpha_b$ is used. This is the integral representation of the Airy function \cite{Olver}: 
\begin{equation}\label{PsiAiry}
    \Psi(z)\propto e^{-i\delta z}\operatorname{Ai}\left(e^{-2 i \pi/3}2^{1/3}\delta^{2/3}(z-1)\right),
\end{equation}
where the integration contour in \eqref{psiint2} is chosen in a standard way \cite{Olver}. 

The asymptotic analysis of the Airy function for \mbox{$|\delta|\gg 1$} proceeds by inspecting stationary points of the exponent in 
\eqref{psiint2}. Clearly there are two such stationary points, $y=\pm\sqrt{2(1-z)}$, corresponding to two components of the WKB solution. As is well established, the coefficients that appear in front of these solutions may exhibit discontinuous jumps upon variation of the argument of the complex variable $z-1$. This discontinuous jump is known as the Stokes phenomenon \cite{Olver} and occurs upon crossing  Stokes lines. Along the Stokes lines, the two WKB solutions differ most significantly, but their phases \eqref{phiMapp}-\eqref{phiPapp} coincide:
\begin{equation}\label{ReCond}
    \text{Re}\{\delta \,\phi_{0}^{(-)}(z)\}=\text{Re} \{\delta\,\phi_{0}^{(+)}(z)\}.
\end{equation}
Near the right turning point, one can use \eqref{SolZ} and rewrite \eqref{ReCond}: 
\begin{equation}
    \text{Re}\{\delta(1-z)^{3/2}\}=0, 
\end{equation}
which gives three Stokes lines emanating from $z=1$ along the angles $-2\arg(\delta)/3$, $2\pi/3-2\arg(\delta)/3$, and $-2\pi/3-2\arg(\delta)/3$ (red lines in Fig. \ref{StokesBrP}). The regions between Stokes lines are called Stokes sectors. The coefficients in front of the WKB solutions in the asymptotic expansion stay fixed inside these sectors and  change only upon crossing Stokes lines.

From the Airy function \eqref{PsiAiry}, one can find that the asymptotic expansion of $\Psi(z)$ contains two WKB solutions inside a single Stokes sector (depicted in green in Fig. \ref{StokesBrP}):
\begin{equation}
    \Psi(z)\propto\Psi_-(z)+\Psi_+(z),
\end{equation}
and one WKB solution inside the remaining two sectors (depicted in blue in Fig. \ref{StokesBrP}):
\begin{equation}
    \Psi(z)\propto\Psi_-(z).
\end{equation}
Thus, the two Stokes lines at angles $-2\pi/3-2\arg(\delta)/3$ and $-2\arg(\delta)/3$ prevent the singular solution $\Psi_+$ from penetrating the region where it diverges.

\section{Keldysh path integral}
\label{AppKpi}

In this appendix, we discuss the Keldysh path integral and properties of the Lindbladian density functional. To begin with, let us consider the Lindblad equation for the reduced density matrix $\hat \rho$:
\begin{equation}
    \p_t \hat{\rho}=\hat{\mathcal{L}}\hat{\rho}=-i\left[\hat{H},\hat{\rho}\right]+\kappa\left(\hat{L}\hat{\rho} \hat{L}^{\dagger}-\frac{1}{2}\left\{ \hat{L}^{\dagger}\hat{L},\hat{\rho}\right\}_+\right),
\end{equation}
where $\hat{\mathcal{L}}$ is a Liouvillian superoperator, $\hat{H}=\hat{H}(\hat{a},\hat{a}^\dagger)$ is a Hamiltonian, $\hat{L}=\hat{L}(\hat{a},\hat{a}^\dagger)$ is one of the jump operators, and $\kappa$ is a  dissipation rate. 
Proceeding from the Lindblad master equation to the Keldysh functional integral as discussed  in \cite{Kamenev} and \cite{Sieberer2016}, one finds for the evolution superoperator:
\begin{equation}
\int\mathcal{D}[\alpha_{+},\bar\alpha_{+},\alpha_{-},\bar\alpha_{-}]\,\,  e^{- S[\alpha_{+},\bar\alpha_{+},\alpha_{-},\bar\alpha_{-}]},
\end{equation}
where the Keldysh action is: 
\begin{equation}
 S =
 \!\int\! dt
 \big[ 
 \bar{\alpha}_{+}\p_t \alpha_{+}
 -\alpha_{-}\p_t \bar{\alpha}_{-}- \mathcal{L}(\alpha_{+},\bar\alpha_{+},\alpha_{-},\bar\alpha_{-})
 \big],
\end{equation}
and $\alpha_+,\alpha_-,\bar{\alpha}_+,\bar{\alpha}_-$ are the coherent bosonic fields on the Keldysh contour. The Lindbladian density function can be decomposed into Hamiltonian and dissipative parts:
\begin{equation}\label{LinHam}
    \mathcal{L}_{H}=-i( H_+- H_-),
\end{equation}
\begin{equation}\label{LinDiss}
    \mathcal{L}_{D}=\kappa (L_+ \overline{L}_{-}-\tfrac{1}{2}\overline{L}_{+} L_+-\tfrac{1}{2}\overline{L}_{-} L_-),
\end{equation}
where the matrix elements  of the {\it normally ordered} Hamiltonian and jump operator are given by \cite{Sieberer2016,Kamenev}: 
\begin{equation}
\begin{gathered}\label{FuncHamandJ1}
 H_{\pm}=H(\alpha_{\pm},\bar{\alpha}_{\pm}), 
\end{gathered}
\end{equation}
\begin{equation}
\begin{gathered}\label{FuncHamandJ2}
{L}_\pm=L(\alpha_{\pm} ,\bar{\alpha}_{\pm}),\quad\overline{L}_{\pm}={L}^\dagger(\alpha_{\pm},\bar{\alpha}_{\pm}),
\end{gathered}
\end{equation}
which are functions of two complex variables. 
The Keldysh rotation introduces the classical and quantum components of the fields as:
$$\alpha_\pm=\alpha_{cl}\pm\alpha_{q}/2,\qquad \overline{\alpha}_\pm=\overline{\alpha}_{cl}\pm\overline{\alpha}_{q}/2.$$

As an example, the Lindbladian density function for the model  (\ref{Hrwa}), (\ref{eq:jump-operators}) takes the following form:
\begin{equation}\label{Li}
\begin{gathered}
      \mathcal{L}=
      \left\{
      i\tilde{\Delta}^*\alpha_{cl}\bar{\alpha}_{q}
      +G \bar{\alpha}_{cl}\bar{\alpha}_{q}
      -\tilde{\eta}^*
      \left(\alpha_{cl}^2+\tfrac{1}{4}\alpha_{q}^2\right)
      \bar{\alpha}_{cl}\bar{\alpha}_{q}
      -c.c.
      \right\}
     \\
     -\frac{1}{2}
     \left\{
     \bar{\alpha}_{q}\alpha_{q}
     \left[
     -i\tilde{\Delta}^*
     +(2\tilde{\eta}^*+\kappa_{\phi})
     \bar{\alpha}_{cl}\alpha_{cl}
     \right]
     +\kappa_{\phi}\alpha_{q}^2\bar{\alpha}_{cl}^2
     +c.c.
     \right\},
\end{gathered}
\end{equation}
where $\tilde{\Delta}=\Delta-i\kappa$ is a complex detuning  and $\tilde{\eta}=\eta-iU$ is a complex nonlinearity.
Taking variations of the Keldysh action \eqref{action} with respect to the four fields, one obtains the instanton equations of motion:
\begin{widetext}
\begin{equation}\label{EqofM0}
\begin{split}
& \partial_t \alpha_{cl}
=i\tilde{\Delta}^{*}\alpha_{cl}
+G \bar{\alpha}_{cl}
-\tilde{\eta}^{*}
\left(
\bar{\alpha}_{cl} \alpha_{cl}^2
+\tfrac{1}{4}\bar{\alpha}_{cl} \alpha_q^2
\right)
+\tfrac{1}{2}\tilde{\eta}\bar{\alpha}_q \alpha_q \alpha_{cl}
-\alpha_q
\left(
\kappa+2\eta \bar{\alpha}_{cl}\alpha_{cl}
\right)
-\kappa_{\phi}\alpha_{cl}
\left(
\alpha_{cl}\bar{\alpha}_{q}
+\alpha_{q}\bar{\alpha}_{cl}
\right),
\\
& \partial_t \bar{\alpha}_{cl}
=-i\tilde{\Delta}\bar{\alpha}_{cl}
+G \alpha_{cl}
-\tilde{\eta}
\left(
\bar{\alpha}_{cl}^2 \alpha_{cl}
+\tfrac{1}{4}\bar{\alpha}_{q}^2 \alpha_{cl}
\right)
+\tfrac{1}{2}\tilde{\eta}^* \alpha_q \bar{\alpha}_q \bar{\alpha}_{cl}
+\bar{\alpha}_q
\left(
\kappa+2\eta \bar{\alpha}_{cl}\alpha_{cl}
\right)
+\kappa_{\phi}\bar{\alpha}_{cl}
\left(
\bar{\alpha}_{cl}\alpha_{q}
+\bar{\alpha}_{q}\alpha_{cl}
\right),
\\
& \partial_t \alpha_q
=i\tilde{\Delta}\alpha_{q}
+G \bar{\alpha}_q
-\tilde{\eta}^*
\left(
\bar{\alpha}_q \alpha_{cl}^2
+\tfrac{1}{4}\bar{\alpha}_q \alpha_q^2
\right)
+2\tilde{\eta} \bar{\alpha}_{cl} \alpha_q \alpha_{cl}
-2\eta \bar{\alpha}_q \alpha_{cl} \alpha_q
-\kappa_{\phi}\alpha_{q}
\left(
\alpha_{cl}\bar{\alpha}_{q}
+\alpha_{q}\bar{\alpha}_{cl}
\right),
\\
& \partial_t \bar{\alpha}_q
=-i\tilde{\Delta}^*\bar{\alpha}_q
+G\alpha_q
-\tilde{\eta}
\left(
\alpha_q \bar{\alpha}_{cl}^2
+\tfrac{1}{4}\alpha_q \bar{\alpha}_q^2
\right)
+2\tilde{\eta}^* \alpha_{cl} \bar{\alpha}_q \bar{\alpha}_{cl}
+2\eta \alpha_q \bar{\alpha}_{cl} \bar{\alpha}_q
+\kappa_{\phi}\bar{\alpha}_{q}
\left(
\alpha_{cl}\bar{\alpha}_{q}
+\alpha_{q}\bar{\alpha}_{cl}
\right).
\end{split}
\end{equation}
\end{widetext}
These equations are required for calculating the instanton trajectories. However, in the variables
$\alpha_{cl}$, $\bar{\alpha}_{cl}$, $\alpha_q$, and $\bar{\alpha}_q$, they do not possess a simple complex-conjugation structure. In other words, the equation for $\bar{\alpha}_{cl}$ is not the complex conjugate of the equation for $\alpha_{cl}$, and the equation for $\bar{\alpha}_q$ is not the complex conjugate of the equation for $\alpha_q$.

To restore the complex conjugation structure, one introduces new variables \cite{Kamenev,Thompson2022}:
\begin{equation}\label{ac}
     \alpha=\alpha_{cl},
     \quad
     \chi=\bar{\alpha}_q,
     \quad
     \bar{\alpha}=\bar{\alpha}_{cl},
     \quad
     \bar{\chi}=-\alpha_q.
\end{equation}
The equations of motion in these variables are presented in the main text, see Eq.~\eqref{EqofM1}. If the Lindbladian density  is self-conjugate:
\begin{equation}\label{scL}
\overline{\mathcal{L}(\alpha,\chi,\bar{\alpha},\bar{\chi})}
=
\mathcal{L}(\alpha,\chi,\bar{\alpha},\bar{\chi}),
\end{equation}
then the equations of motion \eqref{EqofM1} for bar-variables are  complex conjugates of those for non-bar ones. To prove this statement, one needs to discuss the properties of the functions \eqref{FuncHamandJ1}-\eqref{FuncHamandJ2} in new variables:
\begin{equation}
\begin{gathered}\label{FuncHamandJ3}
 H_{\pm}=H(\alpha\mp\bar \chi/2,\bar\alpha\pm\bar\chi/2),
 \\{L}_\pm=L(\alpha\mp\bar \chi/2,\bar\alpha\pm\bar\chi/2),
 \\\overline{L}_{\pm}=\overline{L}(\alpha\mp\bar \chi/2,\bar\alpha\pm\bar\chi/2).
\end{gathered}
\end{equation}
In Eq. \eqref{FuncHamandJ2} functions depend on two  complex variables $a=\alpha\mp\bar \chi/2$ and $b=\bar\alpha\pm\bar\chi/2$. They satisfy the following complex-conjugation properties:
\begin{equation}\label{prop}
\begin{aligned}
&\overline{H(a,b)}=H(\bar b,\bar a),
\\
&\overline{L(a,b)}=\overline L(\bar b,\bar a),\quad
\overline{\overline L(a,b)}=L(\bar b,\bar a).
\end{aligned}
\end{equation}
The first relation reflects that the Hamiltonian is self-adjoint, whereas the last two express connection between the jump operator and its complex-conjugate counterpart. Combining \eqref{prop} with \eqref{FuncHamandJ2} gives the following: 
\begin{equation}\label{ConjFun}
   \overline{H}_{\pm}=H_{\mp},\quad \overline{L_\pm}=\overline{L}_\mp,\quad\overline{\overline{L}_\pm}=L_\mp.
\end{equation}
It follows directly from these properties that both the Hamiltonian part $\mathcal{L}_H$ and the dissipative part $\mathcal{L}_D$ are self-conjugate. As a result, the identity \eqref{scL} is proven, resulting in the self-conjugate equations of motion \eqref{EqofM1}.

\bibliographystyle{apsrev4-1}
\bibliography{main}

\end{document}